\documentclass[fleqn,usenatbib]{mnras}
\pdfoutput=1

\usepackage{txfonts}

\usepackage{threeparttable}
\usepackage{comment}
\usepackage{pdflscape}

\usepackage[T1]{fontenc}
\usepackage{ae,aecompl}

\usepackage{graphicx}	
\usepackage{amsmath}	
\usepackage{amssymb}	

\usepackage{relsize}	

\title[Retired A Stars. A massive problem.]{The masses of retired A stars with asteroseismology: \emph{Kepler} and K2 observations of exoplanet hosts}

\author[T. S. H. North et al.]{
Thomas S. H. North,$^{1,2}$\thanks{E-mail: txn016@bison.ph.bham.ac.uk  (TSHN)}
Tiago L. Campante,$^{3,4,1,2}$
Andrea Miglio,$^{1,2}$
\newauthor
Guy R. Davies,$^{1,2}$
Samuel K. Grunblatt,$^{5}$
Daniel Huber,$^{5,6,7,2}$
\newauthor
James S. Kuszlewicz,$^{1,2}$
Mikkel N. Lund,$^{1,2}$
\newauthor
Benjamin F. Cooke$^{1,2}$
William J. Chaplin,$^{1,2}$
\\
$^{1}$School of Physics and Astronomy, University of Birmingham, Birmingham, B15 2TT, United Kingdom\\
$^{2}$Stellar Astrophysics Centre (SAC), Department of Physics and Astronomy, Aarhus University,\\ Ny Munkegade 120, DK-8000 Aarhus C, Denmark\\
$^{3}$Instituto de Astrof{\' i}sica e Ci{\^e}ncias do Espa{\c c}o, Universidade do Porto, CAUP, Rua das Estrelas, PT4150-762 Porto, Portugal \\
$^{4}$Departamento de F{\' i}sica e Astronomia, Faculdade de Ci\^{e}ncias, Universidade do Porto, Rua do Campo Alegre 687, PT4169-007 Porto, Portugal\\
$^{5}$Institute for Astronomy, University of Hawai`i, 2680 Woodlawn Drive, Honolulu, HI 96822\\
$^{6}$Sydney Institute for Astronomy (SIfA), School of Physics, University of Sydney, NSW 2006, Australia\\
$^{7}$SETI Institute, 189 Bernardo Avenue, Mountain View, CA 94043, USA\\
}

\date{Accepted XXX. Received YYY; in original form ZZZ}

\pubyear{2015}

\begin{document}
\label{firstpage}
\pagerange{\pageref{firstpage}--\pageref{lastpage}}
\maketitle

\begin{abstract}
We investigate the masses of ``retired A stars'' using asteroseismic detections on seven low-luminosity red-giant and sub-giant stars observed by the NASA Kepler and K2 Missions. Our aim is to explore whether masses derived from spectroscopy and isochrone fitting may have been systematically overestimated. Our targets have all previously been subject to long term radial velocity observations to detect orbiting bodies, and satisfy the criteria used by Johnson et al. (2006) to select survey stars that may have had A-type (or early F-type) main-sequence progenitors. The sample actually spans a somewhat wider range in mass, from $\approx 1\,\rm M_{\odot}$ up to $\approx 1.7\,\rm M_{\odot}$. Whilst for five of the seven stars the reported discovery mass from spectroscopy exceeds the mass estimated using asteroseismology, there is no strong evidence for a significant, systematic bias across the sample. Moreover, comparisons with other masses from the literature show that the absolute scale of any differences is highly sensitive to the chosen reference literature mass, with the scatter between different literature masses significantly larger than reported error bars. We find that any mass difference can be explained through use of differing constraints during the recovery process. We also conclude that underestimated uncertainties on the input parameters can significantly bias the recovered stellar masses, which may have contributed to the controversy on the mass scale for retired A stars.

\end{abstract}




\section{Introduction}\label{sec:intro}
Long term radial velocity surveys have discovered a population of giant planets on ${\geq}300$ day orbits around evolved stars that are more massive than the Sun \citep{Johnson2007a,Bowler2010,2011Witten}. These host stars would have been spectral type A on the main sequence. Evolved stars were targeted since A-type stars are hostile to radial velocity observations on the main sequence, due to rapid rotation broadening spectral lines \citep{Johnson2007a}. These stars show a population of planets distinct from the planets discovered via transit surveys, particularly the vast numbers of planets discovered by the NASA \emph{Kepler} and K2 missions \citep{Johnson2010, Borucki2010, 2013Fressin, Howell2014}. It remains unclear if the different populations observed are a single population observed with strong selection effects, or if the different populations of planets truly indicate separate planet formation mechanisms \citep{2005Fish,2010Howard,2015Becker}. 

Recently the masses of evolved stars have been brought into question on several grounds \citep{2011Lloyd,2013Schlaufman}, with the possibility raised that the masses of evolved hosts have been overestimated when derived from spectroscopic observations. The mass of these stars is typically recovered by interpolating grids of stellar models to the observed $T_{\textrm{eff}}$, $\log{g}$, and [Fe/H], and including additional parameters such as luminosity and colours where available (see \citealt{Johnson2007a} and references therein). These stellar models are then explored in a probabilistic fashion to find the best solution for the fundamental stellar properties \citep{2006dasilva,2015Ghezzi}.

These evolved stars have been termed ``retired A stars'' in current literature \citep{Johnson2008,Bowler2010,2011Lloyd}, since the derived masses for these stars is typically $M\gtrsim 1.6\textrm{M}_{\odot}$ i.e around the boundary in stellar mass between A and F type stars on the main sequence.  We follow that convention in this work, but note that the term ``retired A stars'' can extend to the stellar mass range more typically associated with hot F type stars on the main sequence (${\sim}1.3-1.6\textrm{M}_{\odot}$). 

To try and resolve the above issues another analysis method to determine the masses of evolved stars is needed. The high quality data from the \emph{Kepler} and K2 missions provide an opportunity to perform asteroseismology \citep{2010Gill,Chaplin2015_K2C1} on known evolved exoplanet hosts \citep{2017arXiv170401794C}. In this paper we investigate 7 stars that have been labelled ``retired A stars'' in the literature, and use a homogeneous asteroseismic analysis method to provide accurate and precise masses. For the ensemble, we investigate the fundamental stellar properties estimated from differing combinations of spectroscopic and asteroseismic parameters. The stellar masses are estimated by fitting grids of stellar models to the observable constraints. With these masses we address any potential systematic bias in the masses of evolved hosts, when the masses are derived from purely spectroscopic parameters. We also investigate potential biases due to the choice of the stellar models used.

The format of the paper is as follows. Sec \ref{sec:tgt} describes how the targets were selected and vetted. Sec \ref{sec:anal} discusses how the lightcurves were processed to allow the solar-like oscillations to be detected and how the asteroseismic parameters were extracted from the observations, whilst Sec \ref{sec:astero} details any previous mass results for each star in turn, and any subtleties required during the extraction of the asteroseismic parameters. The modelling of the stars to estimate the fundamental stellar properties is discussed in Sec \ref{sec:mod}. The final results are in Sec \ref{sec:results}. In Sec \ref{sec:diss} we explore in detail potential sources of biases in recovering the fundamental parameters, along with a detailed discussion of potential biases induced in stellar modelling due to differences in constraints and underlying physics.

\section{Target Selection}\label{sec:tgt}
Targets were selected from cross-referencing the K2 Ecliptic Plane Input Catalog (EPIC) \citep{2016Huber} and the NASA Exoplanet Archive \citep{ExArc2013}, where only confirmed planets discovered by radial velocity were retained. The resulting list was then cross-checked with the K2FOV tool \citep{k2fov} to ensure the stars were observed during the K2 mission. To ensure these hosts were all selected from the correct area of parameter space, it was also checked that they all passed the target selection of \cite{Johnson2006} of, $0.5<M_{V}< 3.5$, $0.55<B-V<1.0$ \footnote{The selection function also contains an apparent magnitude cut of $V\leq7.6$. We ignore this cut, as this was imposed originally to limit the required exposure time for the stellar spectra and does not influence the fundamental properties of the stars themselves.}. This produces 6 stars in Campaigns 1-10 (C1-10).

The lightcurve for the star identified in C1 was found to be of too low quality to observe stellar oscillations. An additional target was found in C2, through checking targets in K2 guest observer programs\footnote{Targets found using GO programs and targets listed here, https://keplerscience.arc.nasa.gov/k2-approved-programs.html}  of bright evolved stars that have been subject to long term radial velocity observations \citep{2011Witten}. This star was not identified in the initial selection as it is not a host star but it passes the color and absolute magnitude selection of \cite{Johnson2006}.

HD 212771 was also subject to asteroseismic analysis in \cite{2017arXiv170401794C}, using the same methods presented in Sec \ref{sec:mod}.
 
In addition the retired A star HD 185351, observed during the nominal \emph{Kepler} mission, has been added to the sample. This star has already been subject to asteroseismic analysis in \cite{2014JohnsonHuber}. However it has been added to this sample for reanalysis for completeness.

The 7 stars in our ensemble are summarised in Table \ref{tab:Stars}, including which guest observer program(s) the stars were part of. Before we discuss the previous mass estimates for each star in Sec \ref{sec:astero}, we discuss the data collection and preparation required to extract the asteroseismic parameters from the K2 data.

\begin{table*}
	\centering
	\caption{The 7 stars to be investigated in the paper, all have been observed by either the \emph{Kepler} or K2 missions, and subject to long term radial velocity observations. The Obs column indicates what observing campaign of K2 the star was observed in (C2-10), or if it was observed in the \emph{Kepler} mission (KIC). The GO column indicates which K2 guest observer program(s) the star was part of.}
	\label{tab:Stars}
	\begin{tabular}{lllllll} 
		\hline
		EPIC/KIC & HD & Obs & Mag ($V$) & RA (h:m:s) & Dec (d:m:s) & GO \\
		\hline
		203514293 & 145428 & C2*  & 7.75  & 16:11:51.250 & -25:53:00.86 & 2025,2071,2109\\
		220548055 & 4313   & C8  &	7.82  & 00 45 40.359 &	+07 50 42.07 & 8031,8036,8040,8063\\
		215745876 & 181342 & C7  & 7.55  & 19:21:04.233 & -23:37:10.45 & 7041,7075,7084\\
		220222356 & 5319   & C8*  &	8.05  & 00:55:01.400 & +00:47:22.40 & 8002,8036,8040\\
		8566020   & 185351 & KIC* & 5.169  & 19:36:37.975 & +44:41:41.77 & N/A\\
		205924248 & 212771 & C3*  & 7.60  & 22:27:03.071 & -17:15:49.16 & 3025,3095,3110\\
		228737206 & 106270 & C10* & 7.58 & 12 13 37.285 & 	-09 30 48.17 & 10002,10031,10040,10051,10077\\
		\hline
		\multicolumn{4}{l}{*Observed in short cadence mode }\\
		\hline
	\end{tabular}
\end{table*}

\section{Observations and data preparation}\label{sec:anal}
All targets have been subject to long term radial velocity programs attempting to detect the periodic stellar radial velocity shifts induced by orbiting planets. However for the purposes of asteroseismology high quality, uninterrupted photometry is required. This was achieved during the \emph{Kepler} and K2 missions. 

The lightcurves for the K2 targets were produced from the target pixel files using the K2P$^{2}$ pipeline, \citep{K2P2}, and then subsequently corrected using the KASOC filter \citep{kasocfilt}. Table \ref{tab:Stars} indicates if the stars were observed at a cadence of $\sim$1 minute (short cadence) or $\sim$30 minutes (long cadence). 

\begin{figure*}
\includegraphics[scale=0.45]{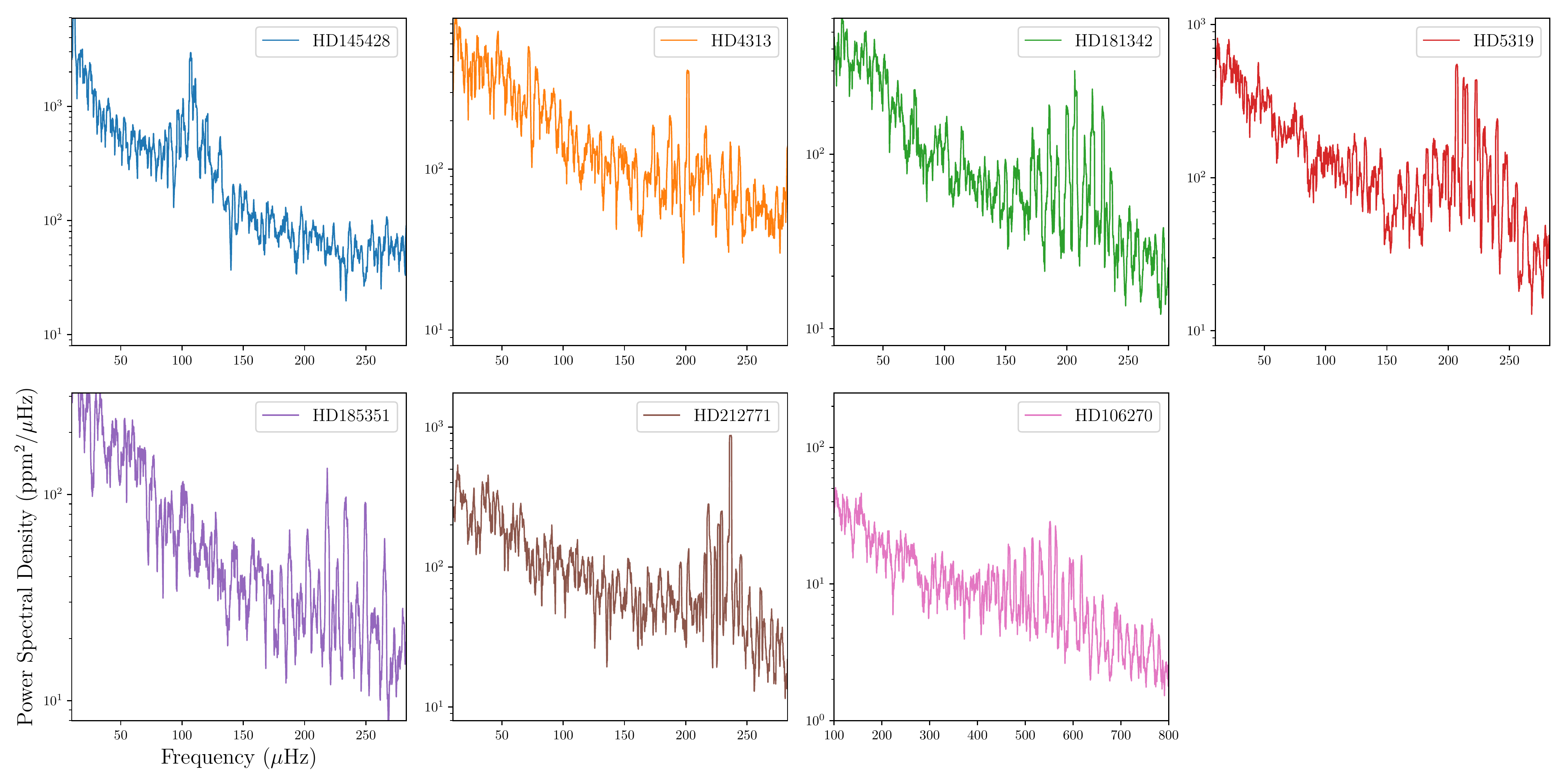}
\caption{The power spectra for each star in our sample, smoothed by a $2\mu$Hz uniform filter ($4\mu$Hz in case of HD 106270), from which we extract the asteroseismic parameters. The stellar oscillations are clearly visible above the granulation background. Note the change in scale for the Campaign 10 star, HD 106270. The stars are presented in order of increasing $\nu_{\textrm{max}}$.}
\label{fig:allpsd}
\end{figure*}

The evolved stars in this paper are expected to exhibit solar-like oscillations, with near surface convection driving global oscillation modes ($p$ and $g$ modes) inside the star. Such oscillations have been observed in thousands of red giants by the \emph{Kepler} and K2 missions \citep{2010Huber,2011Hekker,Stello2013,2015Stello}. Fig \ref{fig:allpsd} shows all the power spectra produced from the corrected lightcurves for the ensemble. In all targets there are clear signatures of solar-like oscillations, above the granulation background.  

Here, we make use of the so-called ``global'' asteroseismic parameters; $\nu_{\textrm{max}}$, the frequency of maximum power and $\Delta\nu$, the average large frequency separation, defined as the average frequency spacing between acoustic oscillation modes of the same angular degree $l$ and consecutive radial order $n$. Table \ref{tab:Stars} is ordered by increasing $\nu_{\textrm{max}}$, as are Tables \ref{tab:res} and \ref{tab:mass_comparison}.

These seismic parameters were extracted from each power spectrum using a variety of well established, and thoroughly tested automated methods \citep{Huber2009,Verner2011,2016Lund_hyades,2016Davies}. The values used in subsequent analysis are those returned by the method described in \cite{Huber2009}. Since multiple pipelines were used to extract the parameters, the uncertainties used in the modelling are the formal errors returned by the \cite{Huber2009} pipeline with the standard deviation of the errors returned from the other methods added in quadrature. This additional uncertainty should account for any unknown systematics in each of the recovery methods. When compared to the seismic values returned by the \cite{Huber2009} pipeline, none of the methods differ by more than $1.3\sigma$ in $\Delta\nu$, and less than $1\sigma$ in $\nu_{\textrm{max}}$. Line-of-sight velocity effects are negligible and do not affect the seismic results \citep{2014Davies}.

An additional asteroseismic parameter, where available, is the average $g$-mode period spacing, accessed through $l=1$ ``mixed'' modes \citep{2011Beck, 2011MosserCorot}. Mixed modes can be highly informative in constraining stellar models, and the core conditions of evolved stars \citep{2011Bedding,2016Lagarde}. Unfortunately due to the shorter length of K2 datasets and hence limited frequency resolution, the period spacing is inaccessible for the 6 K2 targets in our ensemble.

\section{Star-by-Star vetting}\label{sec:astero}
In this section we discuss any individual peculiarities of each star separately. Particular focus is placed on HD 185351, which has been subjected to a suite of investigations throughout and after the nominal \emph{Kepler} mission \citep{2014JohnsonHuber, HD185351_2015, HD185351_2016}. 
All available literature masses for the stars in our ensemble are summarised in Table \ref{tab:litmass} in the appendix. The final seismic and spectroscopic values used in the stellar modelling are summarised in Table \ref{tab:res}. 

\subsection{HD 145428}
The most evolved star in our sample, HD 145428, it is not currently known to host planets, but was a target of the Pan-Pacific Planet Search (PPPS \citealt{2011Witten}) conducted on the Southern sky from the 3.9m Anglo-Australian Telescope. Here we use updated spectroscopic parameters from \citealt{2016Witten}. The target selection for the PPPS is very similar to the target selection used in the Lick \& Keck Doppler survey \citep{Johnson2006, Johnson2007a, Johnson2007b}. This star passes the absolute magnitude selection criteria of \cite{Johnson2006}, however $B-V=1.02$ for this star, slightly over the $B-V\leq1$ selection cut. It was decided to retain this star in the sample despite this. 
Whilst most of the stars in our sample have multiple mass values quoted in the literature, this star appears to have been subject to minimal study, limiting the scope of comparison between asteroseismic and spectroscopic mass estimates. 

\subsection{HD 4313}
\label{sec:supp}
HD 4313, an exoplanet host announced in \cite{Johnson2010}, shows evidence for suppressed $l=1$ modes, first identified as a feature in red giant power-spectra in \cite{Mosser2012}. The cause for such suppression is currently under discussion (see \citealt{2015Fuller,2016NatureStello,2016arXivMosser_dipole}), though in this case we assume that it is not a planet-based interaction, since the planet HD 4313b has an orbital period of approximately 1 year. The limited number of observable oscillation modes also has an impact on the precision of the seismic values, as reflected in the uncertainty on $\nu_{\textrm{max}}$ in Table \ref{tab:res}. 

\subsection{HD 181342}
HD 181342, an exoplanet host reported in \cite{johnson2010b}, has the largest spread in reported masses, with estimates from $1.20-1.89\textrm{M}_{\odot}$ \citep{2016Huber,2016MJones}. 

\subsection{HD 5319}
HD 5319, is the only known multiple planet system in our sample. Both discovery papers list stellar masses in excess of $M>1.5\textrm{M}_{\odot}$ \citep{2007Robinson, 2015Gig}.

\subsection{HD 185351}\label{sec:kicstar}
\defcitealias{2014JohnsonHuber}{J14}
HD 185351 (KIC 8566020), one of the brightest stars in the \emph{Kepler} field, has been monitored as part of a Doppler velocity survey to detect exoplanets \citep{Johnson2006}, though no planet has been found. Additionally in \cite{2014JohnsonHuber} (hereafter \citetalias{2014JohnsonHuber}) the star was studied using asteroseismology, comparing the stellar properties determined from various complementary methods, including an interferometric determination of the stellar radius. 
Several mass values are given, in the range $1.6-1.99\textrm{M}_{\odot}$. As mentioned above, the observed period spacing between mixed modes can be an important constraint on core properties and so global stellar properties. In \citetalias{2014JohnsonHuber}, a period spacing $\Delta\Pi=104.7\pm0.2$s is given. Since we wish to perform a homogeneous analysis for the ensemble, we do not include a period spacing for the star during the recovery of the stellar properties in Sec \ref{sec:mod}.

\subsection{HD 212771}
This is an exoplanet host detected in \cite{johnson2010b}. The mass reported in the discovery paper, $M=1.15\textrm{M}_{\odot}$, is consistent with a retired F or G type star. However, the recent work by \cite{2017arXiv170401794C} provides an asteroseismic mass of $M=1.45\textrm{M}_{\odot}$, promoting this star to being a retired A star. This mass was recovered using the same methodology as used in this work. We present an updated mass in this work, though the shift is negligible. 

\subsection{HD 106270}
The final star in our ensemble, this exoplanet host reported in \cite{2011Johnson} is significantly less evolved than the rest of the ensemble. 

\section{Modelling}\label{sec:mod}
\subsection{Stellar Models}\label{sec:stell_mod}
With the asteroseismic parameters determined for each star, the modelling of the ensemble to extract fundamental stellar properties could now take place. We use \texttt{MESA} models \citep{2011Paxton,2013Paxton} in conjunction with the Bayesian code \texttt{PARAM} \citep{2006dasilva, 2017Rod}. A summary of our selected ``benchmark'' options is as follows; 
\begin{itemize}
\item Heavy element partitioning from \cite{1993Grevesse}.
\item OPAL equation of state \citep{2002Rogers} along with OPAL opacities \citep{1996Iglesias}, with complementary values at low temperatures from \cite{2005Ferguson}.
\item Nuclear reaction rates from NACRE \citep{1999Angulo}.
\item The atmosphere model is taken according to \cite{1966Kris}.
\item The mixing length theory was used to describe convection (a solar-calibrated parameter $\alpha_{\textrm{MLT}} =1.9657$ was adopted).
\item Convective overshooting on the main sequence is set to $\alpha_{\textrm{ov}}=0.2H_{p}$, with $H_{p}$ the pressure scale height at the border of the convective core (more on this in Sec \ref{sec:varyov}). Overshooting was applied according to the \cite{1975Maeder} step function
scheme.
\item No rotational mixing or diffusion is included.
\item When using asteroseismic constraints, the large frequency separation $\Delta\nu$ within the \texttt{MESA} model is calculated from theoretical radial mode frequencies, rather than based on asteroseismic scaling relations.
\end{itemize}

Below we discuss the additional inputs required for the modelling, such as $T_{\textrm{eff}}$, [Fe/H] and luminosity. 

In Sec \ref{sec:varyov} we test the robustness of the asteroseismic masses by varying the underlying model physics, and explore the effects of unaccounted for biases in the stellar observations.

\subsection{Additional modelling inputs}\label{sec:addmod}
In addition to the asteroseismic parameters, a temperature and metallicity value are needed for each star. Since multiple literature values exist for the chosen targets, we had to choose a source for each. To ensure the values are self-consistent, when a literature value was chosen for temperature, we took the stellar metallicity from the same source i.e. matched pairs of temperature and metallicity. To account for unknown systematics additional uncertainties of 59K and 0.062 dex were added in quadrature \citep{Torres2012}, to the chosen literature values. Several of the stars have smaller reported [FeH] error bars than the systematic correction of \cite{Torres2012}, for these stars an error bar of 0.1dex was adopted. 

The stellar luminosity also provides a strong constraint on the modelling. The luminosity may be estimated as follows (e.g. see \citealt{pijpers2003}):
\begin{multline}
\log_{10} \frac{L}{L_{\odot}} = 4.0+
0.4 M_{{\textrm{bol}},\odot} -2.0 \log_{10} {\pi [{\textrm{mas}]}} \\-0.4(V-A_V + BC(V)).
\label{eqn:lum}
\end{multline}
Johnson $V$ magnitudes and uncertainties were taken from the EPIC catalog \citep{2016Huber}, the solar bolometric magnitude $M_{\textrm{bol},\odot}=4.73$ is taken from \cite{Torres2010}, from which we also take the polynomial expression for the bolometric correction\footnote{The polynomial bolometric corrections presented in \cite{Torres2010}, are reprints of values presented in \citet{Flowers1996}, having been corrected for typographical errors in the original} $BC(V)$. 
Finally, the extinction $A_V$ is calculated using \texttt{MWDUST} \citep{mwdust}\footnote{github.com/jobovy/mwdust}, using the 3D dust maps from \citet{Green2015}. 

\subsection{Parallaxes}
Parallaxes $\pi$ were taken from \emph{Hipparcos} \citep{2007HIP}, and the recent Tycho-Gaia Astrometric Solution (TGAS) data release, part of the $Gaia$ Data Release 1 \citep{GAIADR1}. The $Gaia$ parallax is generally preferred. HD 185351 and HD 106270 are both missing $Gaia$ TGAS parallaxes due to their bright apparent magnitudes, with $Gaia$ DR1 missing many stars with $Gaia$ magnitude $G\leq7$. For stars with a TGAS parallax an additional uncertainty of 0.3mas has been added to the formal parallax uncertainty as suggested by \cite{GAIADR1}. \cite{2017arXiv170401794C} previously found that the $Hipparcos$ solution for the distance to HD 212771 is in tension with the asteroseismic solution, whilst the $Gaia$ solution is entirely consistent.

Conversely, the luminosity constructed using Eq \ref{eqn:lum} for HD 145428 was severely discrepant due to the large difference between the $Gaia$ and $Hipparcos$ parallaxes ($5.39\pm0.73$mas and $7.62\pm0.81$mas respectively). When the final stellar radius from the modelling is used along with the input temperature, the constructed luminosity is found to be consistent with the $Hipparcos$ luminosity, but not the $Gaia$ luminosity. As such the $Gaia$ luminosity was also ignored in this case, and all modelling results for this star are reported using a $Hipparcos$ parallax based luminosity. There has been discussion in the literature of possible offsets in the $Gaia$ parallaxes when compared to distances derived from eclipsing binaries \citep{2016stassun} and asteroseismology \citep{2016deridder,2017davies,2017arXiv170504697H}.

\section{Results}\label{sec:results}
Table \ref{tab:res} summarises the asteroseismic and spectroscopic inputs used in the analysis, and the estimated stellar properties returned by \texttt{PARAM} for our benchmark set of chosen input physics. Additional modelling using different constraints and model grids is discussed in Sec \ref{sec:diss}. 

\begin{table*}
	\centering
	\caption{The final asteroseismic and spectroscopic inputs and output stellar parameters from the modelling. The effective temperature and metallicity used for each source are taken as matched pairs from the same source.}
	\label{tab:res}
	\begin{threeparttable}
	\centering
	\begin{tabular}{lllllrllll} 
		\hline
	EPIC/KIC & HD & $\Delta\nu$ ($\mu$Hz)& $\nu_{\textrm{max}}$ ($\mu$Hz) & $T_\textrm{eff}$ (K) & [Fe/H] & &Mass (M$_{\odot}$) &  Radius (R$_{\odot}$) & Age (Gyr)\\
	\hline
	203514293\tnote{*}&145428&10.1 $\pm$ 0.3 &107 $\pm$ 2 &$4818\pm100$\tnote{a}&$-0.32\pm0.12$\tnote{a}&&$0.99^{+0.10}_{-0.07}$&$5.51^{+0.26}_{-0.19}$&$9.03^{+2.79}_{-2.72}$\\\\
	220548055&4313&$14.1\pm0.3$&$201\pm8$&$4966\pm70$\tnote{b}&$0.05\pm0.1$\tnote{b}&&
$1.61^{+0.13}_{-0.12}$&$5.15^{+0.18}_{-0.17}$&$2.03^{+0.64}_{-0.45}$\\\\
	215745876&181342&$14.4\pm0.3$&$209\pm6$&$4965\pm80$\tnote{b}&$0.15\pm0.1$\tnote{b}&&
$1.73^{+0.18}_{-0.13}$&$5.23^{+0.25}_{-0.18}$&$1.69^{+0.47}_{-0.41}$\\\\
	220222356&5319&$15.9\pm0.5$&$216\pm3$&$4869\pm80$\tnote{b}&$0.02\pm0.1$\tnote{b}&&
$1.25^{+0.11}_{-0.10}$&$4.37^{+0.17}_{0.17}$&$5.04^{+1.78}_{-1.30}$\\\\
	8566020\tnote{*} &185351&15.6$\pm$0.2 & 230$\pm$ 7 &$5035\pm80$\tnote{c}&$0.1\pm0.1$\tnote{c}&&$1.77^{+0.08}_{-0.08}$&$5.02^{+0.12}_{-0.11}$&$1.51^{+0.17}_{-0.14}$\\\\
	205924248&212771&16.5$\pm$ 0.3&231 $\pm$ 3 &$5065\pm95$\tnote{d}&$-0.1\pm0.12$\tnote{d}&&
$1.46^{+0.09}_{-0.09}$&$4.53^{+0.13}_{-0.13}$&$2.46^{+0.67}_{-0.50}$\\\\
	228737206\tnote{*}&106270&$32.6\pm0.5$&$539\pm13$&$5601\pm65$\tnote{b}&$0.06\pm0.1$\tnote{b}&&$1.52^{+0.04}_{-0.05}$&$2.95^{+0.04}_{-0.04}$&$2.26^{+0.06}_{-0.05}$\\
		\hline
		\multicolumn{10}{l}{\texttt{PARAM} uses [M/H], which we take to equal [Fe/H] for all stars}\\
		\multicolumn{10}{l}{Quoted errors on mass, radius and age are the $68\%$ credible interval from \texttt{PARAM}}\\

		\hline
	\end{tabular}
	\begin{tablenotes}
	\item[*] Gaia TGAS parallaxes were unavailable, or believed unreliable (see Sec \ref{sec:addmod}), and so the \emph{Hipparcos} parallax is used instead in the construction of the stellar luminosity.
	\item[a] \cite{2016Witten}.
	\item[b] \cite{Mortier2013}.
	\item[c] \cite{HD185351_2015}.
	\item[d] \cite{2017arXiv170401794C}.
	\end{tablenotes}
	\end{threeparttable}
\end{table*}

\begin{figure*}
\includegraphics[width=1.5\columnwidth]{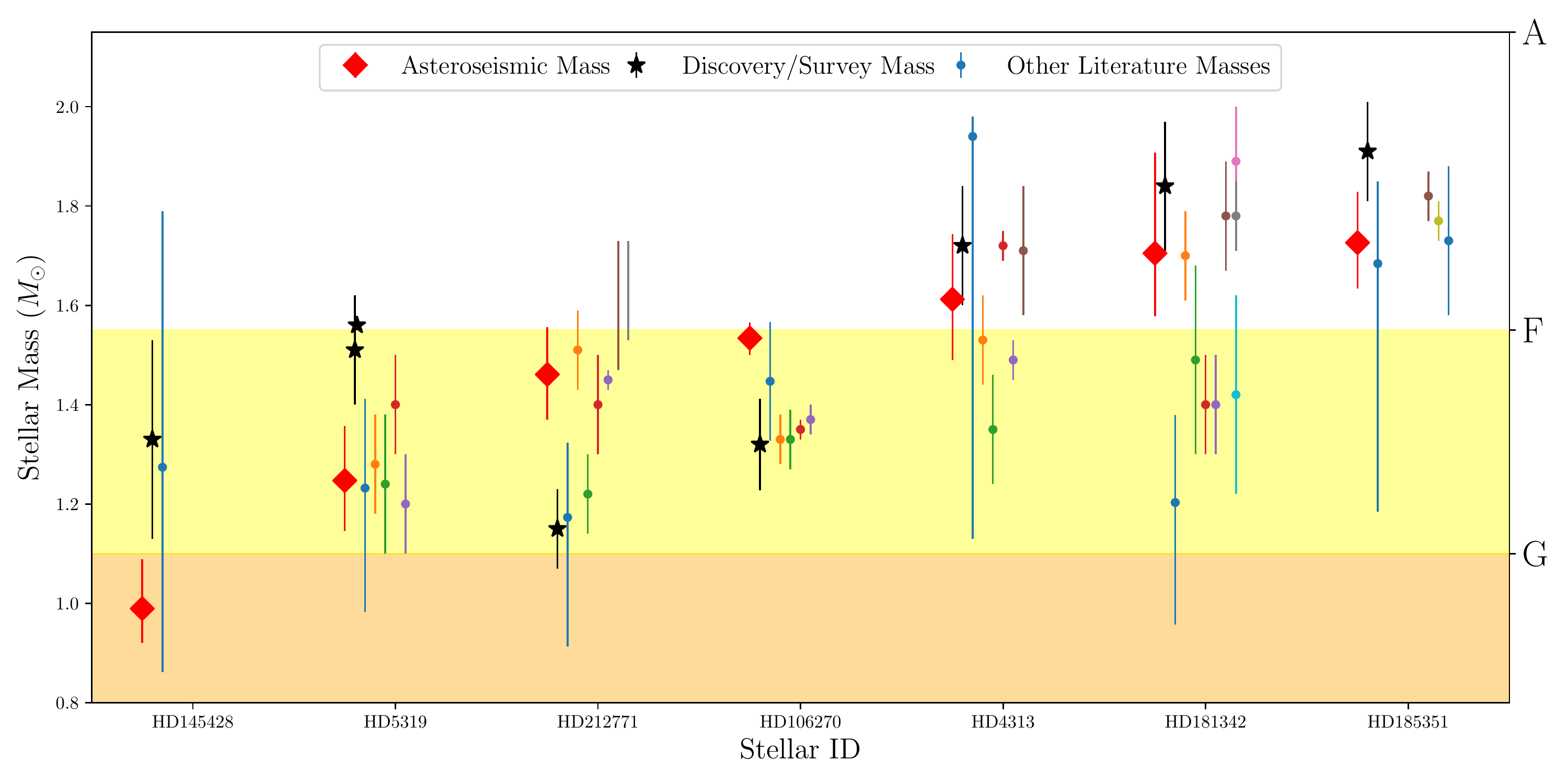}

\caption{Mass vs ID, the horizontal bars indicate approximate spectral type on the main sequence, (AFG corresponding to white, yellow and orange respectively). Black stars indicate the mass of the star as reported in the planet survey or planet detection paper. Red diamonds indicate the \texttt{PARAM} stellar mass from Table \ref{tab:res}, whilst dots indicate other literature values for each star. As can be seen for several of the stars  (HD 5319, HD 145428, HD 181342 and HD 212771), the different mass estimates can cover the entire spectral range of G to A type.}
\label{fig:masses_id}
\end{figure*}

With the results from \texttt{PARAM} we can now compare the stellar masses derived from asteroseismology with the other literature values. Fig \ref{fig:masses_id} shows the different mass estimates from available literature sources (see Table \ref{tab:litmass}), with the masses reported in the planet discovery or survey paper, our primary comparison mass, as black stars. The asteroseismic masses (red diamonds) are shown alongside other literature values (points). 

For the survey mass of HD 185351 no error was provided with the value in \cite{Bowler2010}. We adopt the $\sigma_{M}=0.07$ from \cite{2014JohnsonHuber}, who in their interpolation of spectroscopic parameters onto isochrones recovered a similar mass to \cite{Bowler2010}. 

The survey mass of HD 145428 in \cite{2016Witten} also has no reported formal error bar, although the work quotes ``typical uncertainties 0.15-0.25$\textrm{M}_{\odot}$''. As such we take 0.2$\textrm{M}_{\odot}$ as the uncertainty. 

\begin{figure*}
\includegraphics[width=1.5\columnwidth]{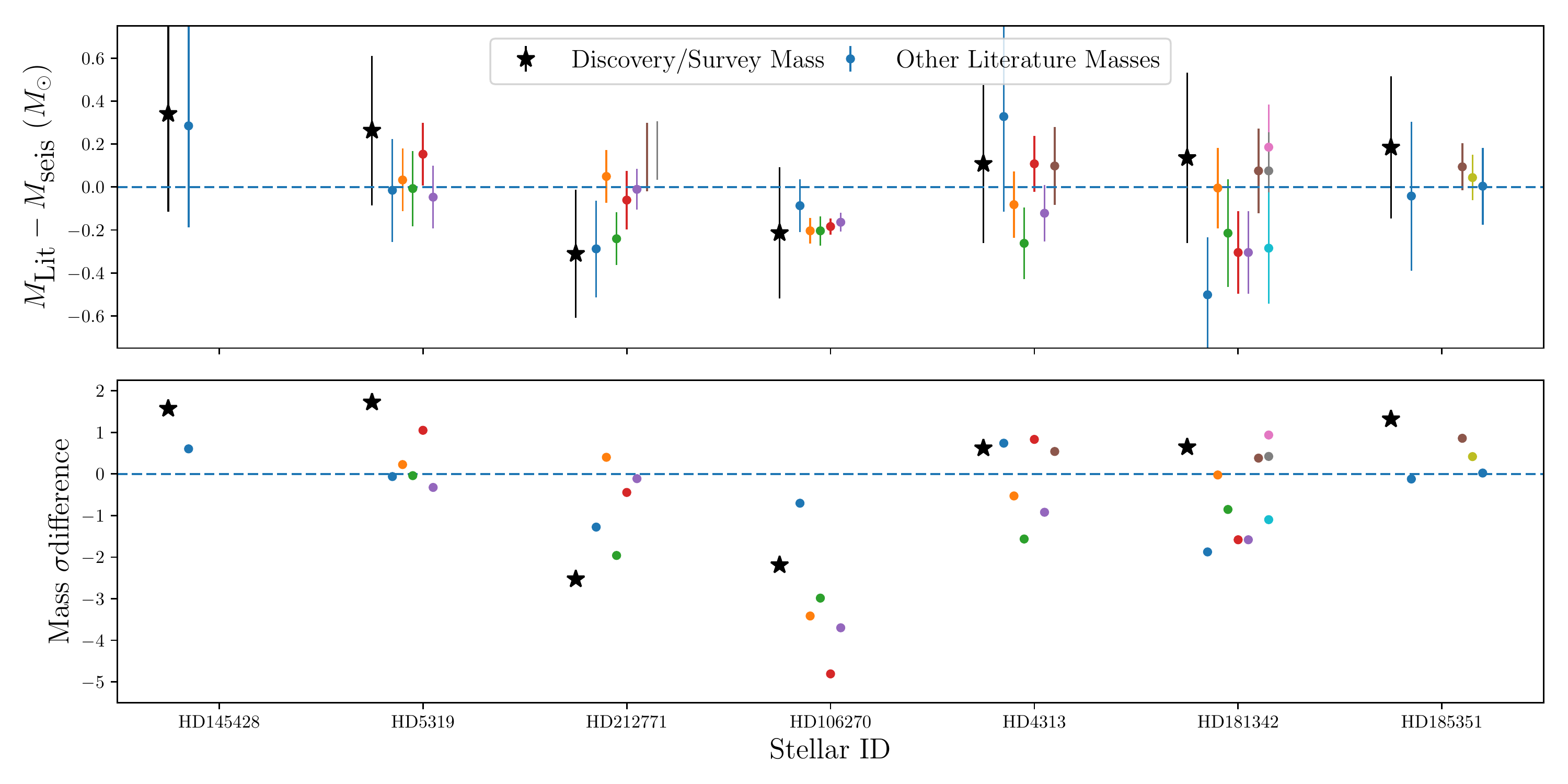}

\caption{Difference between literature and asteroseismic masses, against stellar ID, arranged by increasing stellar mass. Negative values indicate the asteroseismic mass is greater than the literature mass. Again we plot the mass difference with the planet survey mass as a black star. The error bars are the mean seismic error added in quadrature to the literature error bar. The lower panel shows the $\sigma$ difference between the seismic mass and literature mass, where the errors have been added in quadrature. }
\label{fig:masses_diff}
\end{figure*}

To investigate the mass discrepancies between different methods, we plot the difference between the literature values and the asteroseismic masses, versus stellar ID, after arranging stars in the sample in terms of increasing stellar mass, in Fig \ref{fig:masses_diff}. 

A striking feature of Fig \ref{fig:masses_id} is the size of the error bars on each literature mass value, compared to the scatter on the mass values. Several stars have literature mass values with reported error bars of $\approx0.1\textrm{M}_{\odot}$ i.e. quite precise estimates, but with individual mass estimates scattered across a $\geq0.5\textrm{M}_{\odot}$ region. HD 4313, HD 181342 and HD 212771 all show this level of scatter on literature mass values.

Fig \ref{fig:masses_diff} shows the difference in mass estimates, taking the asteroseismic mass as the reference value. Values below zero indicate the literature mass is lower than the seismic mass. The lower panel in the figure shows the difference in standard deviations between the mass estimates, where the literature mass and asteroseismic mass have had their errors added in quadrature. As can be seen, 5 of the 7 stars display asteroseismic masses below the masses reported in the planet discovery/survey paper, however at a $\le2\sigma$ level. HD 212771 and HD 106270 show the opposite behaviour. We caution against taking this difference in asteroseismic masses to be evidence for a systematic shift in stellar mass, due to the small sample size. The average mass offset of the seismic to survey mass is $\Delta\textrm{M}=0.07\pm0.09\textrm{M}_{\odot}$.

A simple Monte-Carlo test was performed to investigate the probability that 5 out of 7 proxy spectroscopic masses would exceed 5 proxy seismic masses for our quoted uncertainties, assuming both of the masses are drawn from normal distributions with a mean of the seismic mass. We found that for a million independent realisations, 16\% of the time 5 of the proxy spectroscopic masses exceed the proxy seismic masses. As such, we see that there is no clear bias between asteroseismic masses and other methods. If we instead derive model independent asteroseismic masses- using the well known asteroseismic scaling relations (e.g. see discussion in \citealt{Chaplin2013})- we find no difference to this result, or other results in the paper.

In Sec \ref{sec:intro} we discuss that the term ``retired A star'' can be used to describe masses associated with hot F stars as well as A type stars. However, if we consider the masses in Table \ref{tab:res}, neither HD 145428 ($0.99\textrm{M}_{\odot}$) or HD 5319 ($1.25\textrm{M}_{\odot}$) can be categorized as such. If we therefore discard these stars, then only 3 of the 5 remaining stars  have seismic masses below the survey mass, with the average offset $\Delta\textrm{M}=0.02\pm0.09\textrm{M}_{\odot}$.

Since the survey masses are a heterogeneous sample of masses, we also compare the seismic masses to several other literature sources (see Table \ref{tab:litmass}). Unfortunately, no single source has masses for all of our stars, and so each homogeneous set of reference masses is a subset of the ensemble. The average ratios are shown in Table \ref{tab:litcomp}. This choice of reference literature mass has a strong impact on the size (and sign) of any observed mass offset. Fig \ref{fig:massratio} shows the distribution of mass ratios for each literature reference mass. 

\begin{table}
	\centering
	\caption{The mean fractional offset of various sets of homogeneous literature mass sources compared to the seismic mass. }
	\label{tab:litcomp}
	\begin{threeparttable}
	\centering
	\begin{tabular}{lll} 
		\hline
		Reference						&$N$ Stars & Offset\\
		\hline
		Discovery/Survey				&7			&$1.07\pm0.07$\\
		\cite{Mortier2013}\tnote{a}		&5			&$0.97\pm0.03$\\
		\cite{Mortier2013}\tnote{b}		&5			&$0.88\pm0.03$\\
		\cite{Jofre2015}				&4			&$1.06\pm0.01$\\
		\cite{2015Bofanti}				&5			&$0.97\pm0.05$\\
		\cite{2016Bofanti}				&5			&$0.91\pm0.03$\\
		\hline
	\end{tabular}
	\begin{tablenotes}
	\item[a] \cite{2013Tsantaki} line list.
	\item[b] \cite{2007Hekker} line list.
	\end{tablenotes}
	\end{threeparttable}
	\end{table}

\section{Discussion}\label{sec:diss}
In order to investigate the robustness of recovering single star masses from stellar models, with or without the inclusion of asteroseismic parameters, we now explore the potential biases from the use of different stellar models, inputs, and error bars on the recovered stellar mass. 

\subsection{Use of different constraints}
Throughout this work we have considered the asteroseismic mass to be the mass returned by \texttt{PARAM} using all available asteroseismic, spectroscopic and parallax/luminosity constraints. To see how much the non-asteroseismic constraints  are influencing the final stellar mass, we also ran \texttt{PARAM} using only $T_{\textrm{eff}}$, [Fe/H] and luminosity, i.e. without seismology. This was done in an effort to emulate the procedure used in \cite{johnson2010b}, using the same constraints, but differing model grids. \cite{2015Ghezzi} previously found that stellar masses can be recovered to good precision using \texttt{PARSEC} models and only spectroscopic constraints. The different mass results for each star are shown in Fig \ref{fig:masses_grids} and are summarised in Table \ref{tab:mass_comparison}. 

Before we discuss the results, we introduce the additional modelling performed using different underlying physics in the stellar models chosen.

\subsection{Use of different model grids}\label{sec:varyov}
To test the sensitivity of the derived stellar masses to the models used, extra grids of \texttt{MESA} models were created. The models described in Sec \ref{sec:mod} include a convective-overshooting parameter $\alpha_{\textrm{ov}}$ during the main sequence, which changes the size of the helium core during the red giant branch phase. This was set to $\alpha_{\textrm{ov}}=0.2H_{p}$ where $H_{p}$ is the pressure scale height. To investigate the impact of changing the underlying model physics on the final stellar mass result, new grids of models with the overshooting parameter adjusted to $\alpha_{\textrm{ov}}=0.1H_{p}$ or 0 were generated and \texttt{PARAM} run using these new models. This parameter was chosen to be varied, since stars in the mass range of retired A stars have convective core during the main sequence. Two further grids of models were used. First, we adopted \texttt{PARSEC} stellar models, which parameterise overshooting as a mass dependent parameter (see \citealt{2012Bressan,2015Bossini}). Second, a grid of \texttt{MIST} (MESA Isochrones and Stellar Tracks, \citealt{2016Choi}) models was used in conjunction with the code \texttt{Isoclassify}.\footnote{see and https://github.com/danxhuber/isoclassify for full details of the code}

Isolating any stellar mass difference between \texttt{MESA} and \texttt{PARSEC} to a single parameter is not possible, since multiple model parameters are different between the models. Additionally there are multiple differences between the \texttt{MESA} tracks used in Sec \ref{sec:mod} and the \texttt{MIST} tracks. The use here of these different tracks is to explore overall mass differences between the different grids, and not to define the precise cause of such a difference.

If we consider first the lower mass stars (HD 5319, HD 145428 and HD 212771), there is no clear trend with overshooting, nor does the inclusion of seismology produce a noticeable shift in mass, with the exception of the \texttt{PARSEC} tracks. The inclusion of luminosity alongside seismic constraints provides the smallest uncertainties.   

For the higher mass stars (HD 4313, HD 181342 and HD 185351), there is a clear trend in increasing mass with decreasing overshooting parameter. The recovered masses using seismic constraints are also in general above the mass estimates without seismic constraints. Whilst for HD 4313 the shift in mass is fairly minor, for  HD 181342 and HD 185351 the mass offset is $\Delta M\sim$0.2$\textrm{M}_{\odot}$. The greatest disparity is between \texttt{PARSEC} results with and without seismic constraints. Again we note that for 5 of the 7 stars in the ensemble, all of the recovered mass estimates are below the masses reported in the planet discovery papers. 

Finally we look at the subgiant HD 106270. There appears to be no strong mass-overshooting parameter dependence, however the \texttt{MESA} models produce significantly different masses to the \texttt{PARSEC} and \texttt{MIST} models that should be investigated more closely. Additionally, the mass estimates recovered from the \texttt{MESA} models without seismic constraints are significantly lower with $\Delta M\sim$0.2$\textrm{M}_{\odot}$.

\begin{figure*}
\includegraphics[width=1.5\columnwidth]{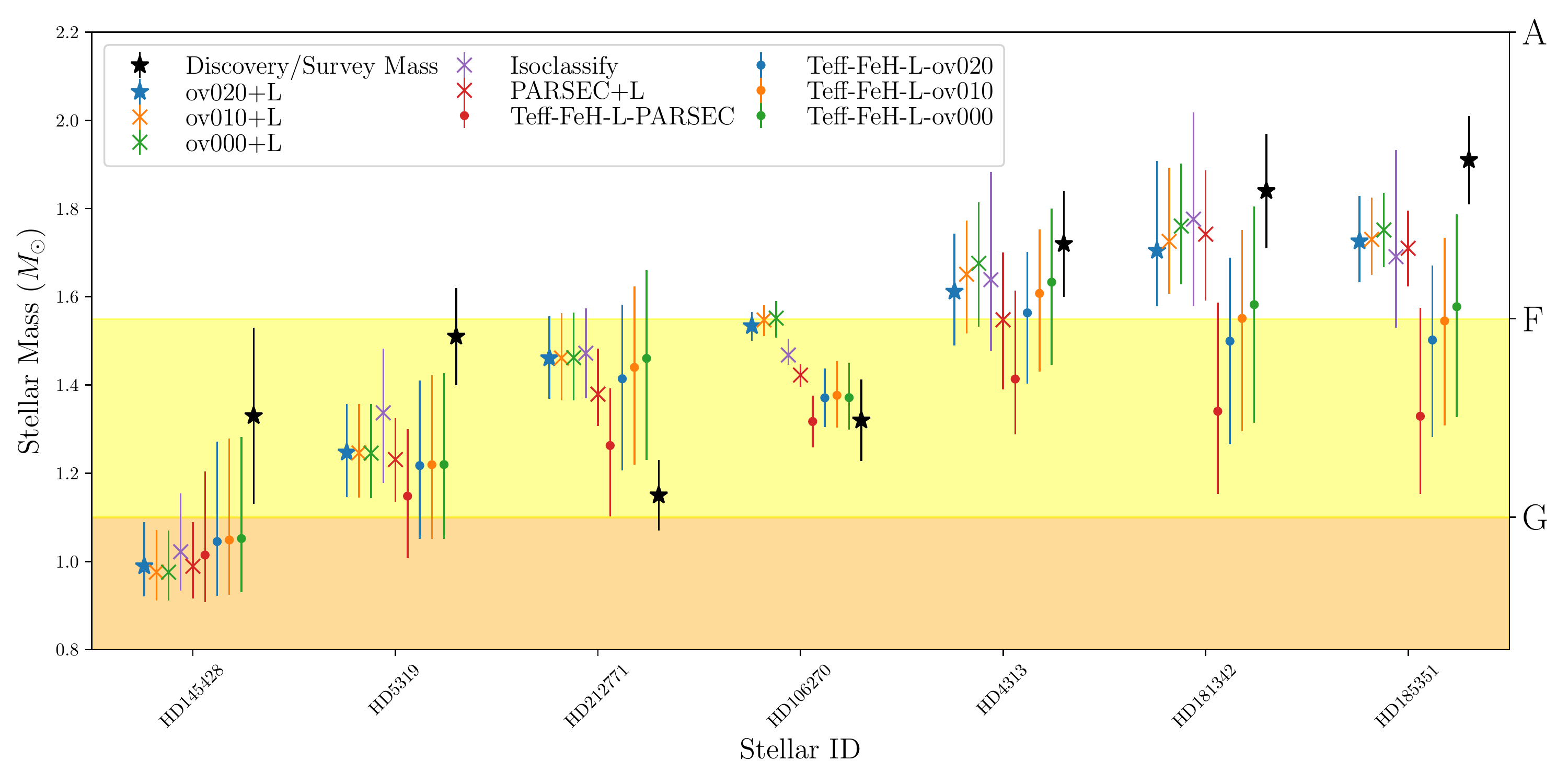}
\caption{All the different mass estimates using \texttt{MESA}, \texttt{PARSEC} and \texttt{MIST} grids (labelled \texttt{Isoclassify}), the \texttt{MESA} and \texttt{PARSEC} grids were ran with and without seismic constraints. Blue stars are the results in Table \ref{tab:res}. Crosses are masses with seismic and luminosity constraints, points are non-seismic constraints only ($T_\textrm{eff}$, [FeH], L). The overshooting parameter used in the models is indicated by the ``ov0*'' label. \texttt{PARSEC} tracks show significant shifts, as do non-seismic results at higher masses. Subgiant HD 106270 shows unique behaviour most likely due to being a subgiant, rather than red giant. }
\label{fig:masses_grids}

\end{figure*}

When we consider the masses returned without the use of the seismology (blue points in Fig \ref{fig:masses_grids}) emulating \cite{johnson2010b}, using the same underlying models as was used with the seismic constraints in Sec \ref{sec:mod}, we fail to recover the same mass as is reported in the discovery paper in most cases. This disparity is presumably due to differences in the underlying stellar models.

When comparing the \texttt{MIST} masses (purple crosses) to the benchmark seismic masses (blue stars), the \texttt{MIST} results typically recover a higher mass. If we include HD 106270, for which the \texttt{MIST} mass is ${\sim}0.1\textrm{M}_{\odot}$ lower than the benchmark mass, then the average mass offset of the \texttt{MIST} masses is $\Delta M=0.03\pm0.06\textrm{M}_{\odot}$. However if we remove HD 106270 the \texttt{MIST} average mass shift is $\Delta M=0.05\pm0.04\textrm{M}_{\odot}$.

\begin{table*}
	\centering
	\caption{Comparing stellar masses estimated using differing physics and constraints. All in Solar masses $\textrm{M}_{\odot}$}
	\label{tab:mass_comparison}
	\begin{threeparttable}
	\centering
\begin{tabular}{llllll}
	\hline
EPIC/KIC& HD & \texttt{MESA} Without Seismology\tnote{a}& \texttt{MESA}\tnote{b} ($\alpha_{\textrm{ov}}=0.0H_{p}$)& \texttt{PARSEC}\tnote{c} & Isoclassify\tnote{d} \\
\hline
203514293&145428&$1.05^{+0.23}_{-0.12}$&$0.97^{+0.09}_{-0.06}$&$0.99^{+0.10}_{-0.07}$&$1.17^{+0.14}_{-0.12}$\\
220548055&4313&$1.56^{+0.14}_{-0.16}$&$1.67^{+0.14}_{-0.14}$&$1.55^{+0.15}_{-0.16}$&$1.67^{+0.18}_{-0.15}$\\
215745876&181342&$1.50^{+0.19}_{-0.23}$&$1.75^{+0.14}_{-0.13}$&$1.74^{+0.14}_{-0.15}$&$1.82^{+0.20}_{-0.20}$\\
220222356&5319&$1.22^{+0.19}_{-0.17}$&$1.24^{+0.11}_{-0.10}$&$1.23^{+0.09}_{-0.10}$&$1.35^{+0.11}_{-0.11}$\\
8566020&185351&$1.50^{+0.17}_{-0.22}$&$1.74^{0.08}_{-0.08}$&$1.80^{+0.09}_{-0.09}$&$1.83^{+0.12}_{-0.10}$\\
205924248&212771&$1.41^{+0.17}_{-0.21}$&$1.45^{+0.10}_{-0.10}$&$1.38^{+0.10}_{-0.07}$&$1.47^{+0.09}_{-0.08}$\\
228737206&106270&$1.37^{+0.07}_{-0.07}$&$1.56^{+0.04}_{-0.04}$&$1.42^{+0.02}_{-0.02}$&$1.41^{+0.03}_{-0.06}$\\
	\hline
	\end{tabular}
	\begin{tablenotes}
	\item[a] \texttt{MESA} models ($\alpha_{\textrm{ov}}=0.2H_{p}$) with [FeH], $T_{\textrm{eff}}$ and luminosity
	\item[b] \texttt{MESA} models ran with $\alpha_{\textrm{ov}}=0.0H_{p}$
	\item[c] \texttt{PARSEC} models 
	\item[d] \texttt{MIST} models ran with \texttt{isoclassify}
	\end{tablenotes}
	\end{threeparttable}
\end{table*}

\subsection{Potential biases in spectroscopic parameters}
\begin{figure}
\includegraphics[width=\columnwidth]{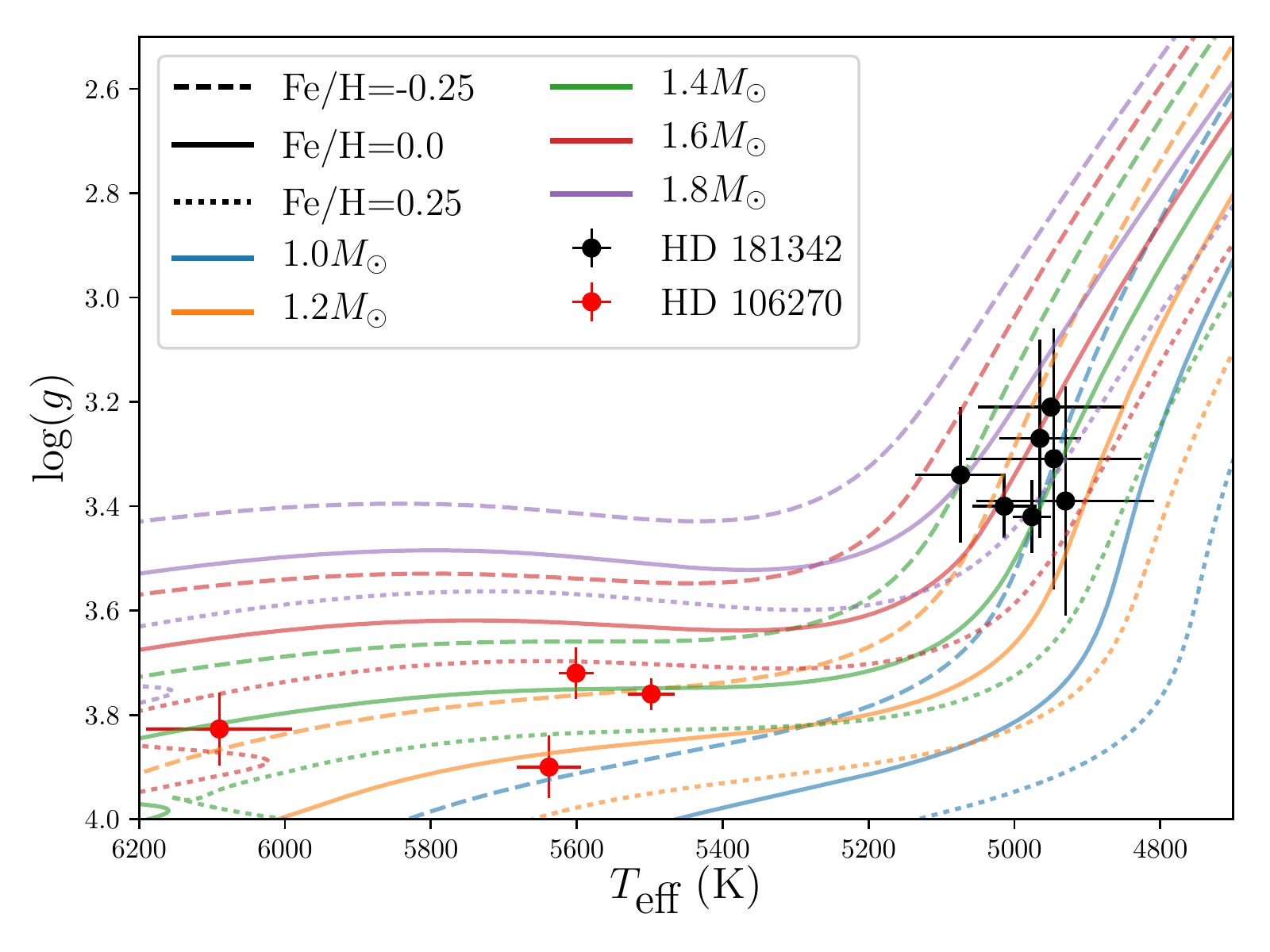}
\caption{Literature spectroscopic parameters for two of the stars in the sample, showing multiple stellar tracks crossed within the uncertainty region. HD 106270 (red) shows that whilst highly precise spectroscopic values are reported in the literature, this limits the parameter space that isochrone fitting can explore, which can lead to disagreement in recovered masses at a significant level. HD 181342 near the base of the red giant branch shows the convergence of the stellar tracks in that region, increasing the difficultly of recovering the stellar parameters.}
\label{fig:mesatracks}
\end{figure}

An additional discrepancy to highlight is that not only are the final mass values derived from spectroscopic parameters in disagreement with each other, possibly caused by differing physics in the models used in each paper during the recovery of the stellar mass, but also the underlying spectroscopic values ($T_{\textrm{eff}}$, $\log{g}$, and [FeH]) can be discrepant at a significant level. Fig \ref{fig:mesatracks} highlights this problem. For two of the stars in the sample, HD 106270 and HD 181342, literature $T_{\textrm{eff}}$ and $\log{g}$ values are plotted over a grid of \texttt{MESA} tracks, using the same physics as in Sec {\ref{sec:mod}}. In particular HD 106270 highlights that reported spectroscopic values may be highly precise, but show significant disagreement to other literature values (see \citealt{2016Blanco} for more discussion on the impact of different spectroscopic pipelines and assumed atmospheric physics on derived parameters). The reported values are highly scattered across the subgiant branch in the Kiel diagram in Fig \ref{fig:mesatracks}. HD 181342, shown in black, highlights the additional problem with targeting red giant branch stars for planet surveys. As the star evolves off the subgiant branch, and begins the ascent of the giant branch, the stellar evolutionary tracks across a wide range of masses converge into a narrow region of parameter space, with tracks of differing mass and metallicity crossing. In the case of HD 181342, taking only the [FeH]=0.0 tracks, masses from $1.2-1.8\textrm{M}_{\odot}$ are crossed. 

In this highly degenerate parameter space it is naturally difficult to search for and isolate the true stellar mass, requiring highly precise and accurate temperatures to help alleviate the degeneracy. It is in this area that the benefits of asteroseismology become clear, as the additional constraint, provided by the asteroseismic observations allow us to break the degeneracy between the spectroscopic parameters to recover a better estimate of the stellar mass.

Whilst the temperature and metallicity uncertainties presented in Table \ref{tab:Stars} are around 0.1 dex and $\sim$80K respectively, several of the planet discovery papers for the stars in the ensemble present much smaller uncertainties, e.g.,  HD 4313, HD 181342, HD 212771 all presented in \cite{johnson2010b} have reported errors $\sigma_{\textrm{[FeH]}}=0.03$dex, $\sigma_{\textrm{Teff}}=44K$ and $\sigma_{L}=0.5\textrm{L}_{\odot}$, as does HD 106270 in \cite{2011Johnson}. \cite{2015Gig} quote the same $\sigma_{\textrm{[FeH]}}$ and $\sigma_{\textrm{Teff}}$ for HD 5319. These spectroscopic parameters and uncertainties were recovered using the package \texttt{Spectroscopy Made Easy} (SME, \citealt{1996Valenti_SME}).

\begin{figure*}
\includegraphics[width=1.5\columnwidth]{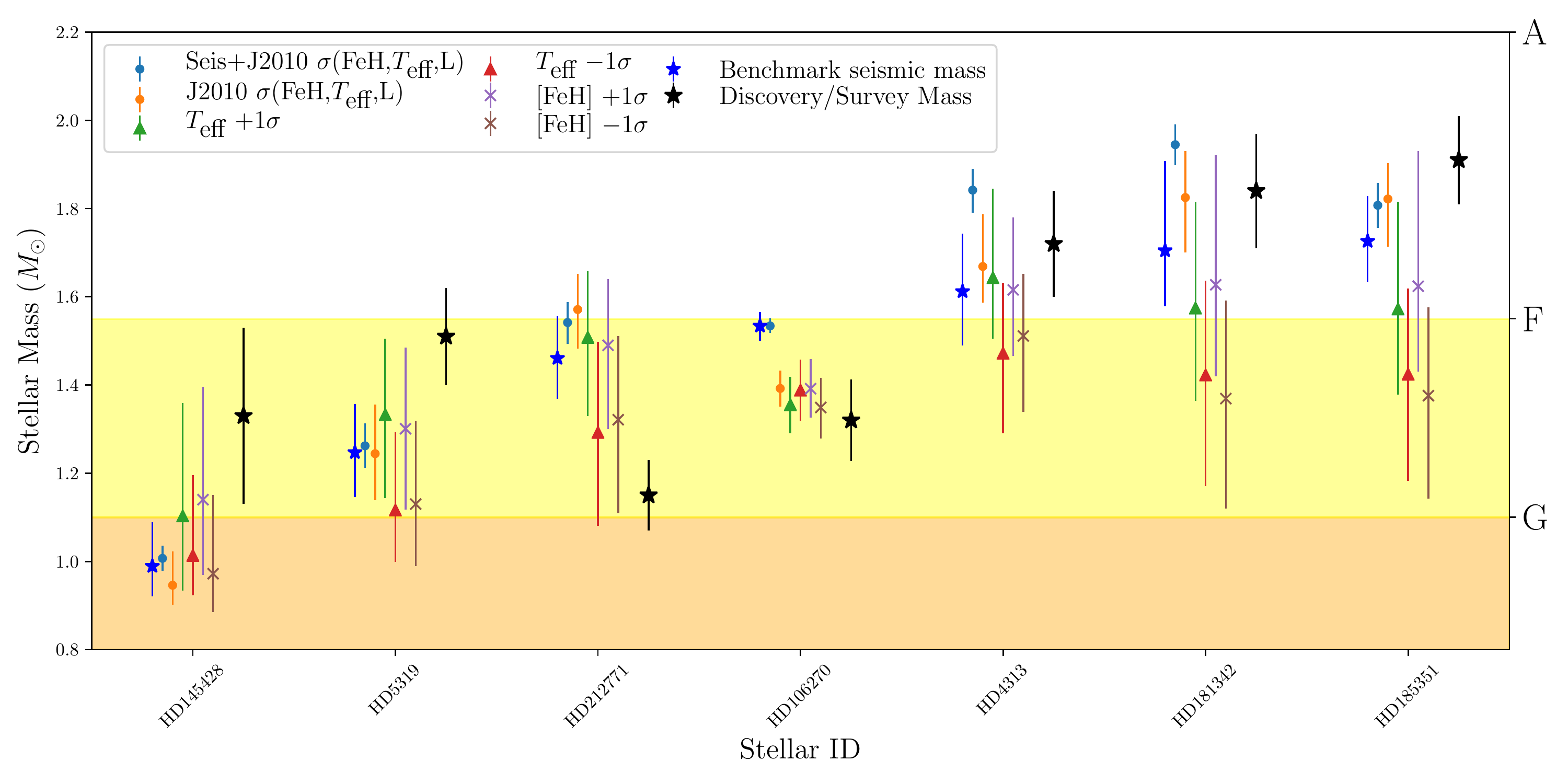}
\caption{Investigating the effect of biases in the spectroscopic parameters, and underestimated error bars. Black stars are the discovery masses, blue stars are the benchmark asteroseismic masses in Table \ref{tab:res}. Blue and orange dots are the same inputs as Table \ref{tab:res} with deflated errors to $\sigma_{\textrm{[FeH]}}=0.03$dex, $\sigma_{T\textrm{eff}}=44$K and $\sigma_{L}=0.5\textrm{L}_{\odot}$, with and without seismic constraint respectively. Remaining markers are the inclusion of $1\sigma$ biases in $T_{\textrm{eff}}$ (triangles) and [FeH] (crosses), using the error values in Table \ref{tab:res}}
\label{fig:masses_bias}
\end{figure*}

To explore what impact such tight error bars might have on inferred stellar masses, \texttt{PARAM} was run once more, using the \texttt{MESA} models with overshooting set to $\alpha_{\textrm{ov}}=0.2H_{p}$ (i.e. identical physics and constraints to the masses in Table \ref{tab:res}), with the inclusion of the asteroseismic constraints in the fitting. The one change here was a systematic reduction of the error bars on [Fe/H], $T_{\textrm{eff}}$ and $L$. In theory, since the same input values and physics are being used, the same values for the stellar mass should be recovered, however this is not what we find. We have effectively shrunk the available parameter space for \texttt{PARAM} to explore. This parameter space is smaller since the error bars on the input parameters define the width of prior used in the Bayesian methodology. With smaller uncertainties the prior is narrower, and so influences the final results more strongly if the underlying value lies away from the mean of the prior (see \citealt{2006dasilva} for more details). To investigate how strongly the seismic values were influencing the recovered parameters, we also ran \texttt{PARAM} using the smaller error bars, without seismic constraints being used, and these results are shown as orange points in Fig \ref{fig:masses_bias}. 
The blue points in Fig \ref{fig:masses_bias} are the results from \texttt{PARAM} using the reduced uncertainties, but including seismic constraints. These mass estimates should agree with the blue stars (the benchmark asteroseismic mass from Table \ref{tab:res}) given the same underlying physics and physical parameters. The only change is a reduction in the size of the error bars on temperature, metallicity and luminosity. What we instead see are significant departures from parity, with generally increasing disagreement as a function of increasing mass (though HD 185351 is an exception). This suggests that potentially inaccurate effective temperatures quoted at high precision can prevent the recovery of the true stellar mass.  
The orange points on Fig \ref{fig:masses_bias} are the results of just using the non-seismic constraints with the deflated error bars discussed above. The recovered mass should be the same as the blue points in Fig \ref{fig:masses_grids} (see Table \ref{tab:mass_comparison} for masses). Instead we see an average offset of $\Delta M\sim$0.1$\textrm{M}_{\odot}$ below the mass in Fig \ref{fig:masses_grids}. This is most likely due to the limited parameter space preventing a full exploration of solutions.

As further tests of the impact of bias in the spectroscopic parameters, \texttt{PARAM} was also run using only the spectroscopic parameters, with artificial biases of $1\sigma$ included on $T_{\textrm{eff}}$ and [FeH]. As Fig \ref{fig:masses_bias} shows, the inclusion of $1\sigma$ shifts in $T_{\textrm{eff}}$ (triangles) or [FeH] (crosses) induces shifts of $\sim$0.2$\textrm{M}_{\odot}$ in stellar mass for the giant stars. 

The subgiant, HD 106270, shows quite separate behaviour and appears more resistant to biases, though it does display a strong disparity in stellar mass estimates with or without asteroseismic constraints. This may be due to the wider separation between tracks of differing mass and metallicity at this point in the HR diagram (as seen in Fig \ref{fig:mesatracks}). Additionally, since most of the evolution is ``sideways'' on the subgiant branch, as the star retains a similar luminosity across a range of temperatures, a single mass track can recover the observed spectroscopic parameters (and luminosity) with an adjustment in stellar age.

One potential issue that has not yet been addressed is the systematic offset in $\log{g}$ between seismology and spectroscopy.  We compared the $\log_{\textrm{seis}}{g}$ recovered with the benchmark seismic mass, to the $\log_{\textrm{spec}}{g}$ reported in the literature source from which we take the $T_{\textrm{eff}}$ and [FeH]. We find the spectroscopic gravities to be overestimated by an average of 0.1dex. Since the spectroscopic parameters are correlated, this may have introduced biases in the temperature and metallicity we have used in the modelling. To test the impact of any bias, we correct the $T_{\textrm{eff}}$ and [FeH] by $\Delta T_{\textrm{eff}}=500\Delta\log{g}$[dex], and $\Delta$[FeH]=$0.3\Delta\log{g}$[dex] (\citealt{Huber2013}, see Figure 2 and surrounding text therein). \texttt{PARAM} was re-run using the \texttt{MESA} models described in Sec \ref{sec:mod}, with the inclusion of the seismic parameters. The mean shift in mass with respect to the benchmark seismic mass was $\Delta M=-0.0097\pm0.010\textrm{M}_{\odot}$. As such, we do not see any evidence for a significant shift in the estimated masses.

\subsection{Potential biases in asteroseismic parameters}
To ensure a thorough test of potential biases, the input $\nu_{\textrm{max}}$, $\Delta\nu$ values were also separately perturbed by $1\sigma$ and \texttt{PARAM} was re-run. We note that in Table \ref{tab:res} both seismic parameters are  given to a similar level of precision as the temperatures (average precision on $\nu_{\textrm{max}}=2.4\%$, $\Delta\nu=2.2\%$ and $T_{\textrm{eff}}=1.6\%$). Each $1\sigma$ perturbation produced an mean absolute shift in mass of $\lesssim0.04\pm0.009\textrm{M}_{\odot}$. These mass shifts are approximately five times smaller than those given by the $1\sigma$ perturbations to the spectroscopic parameters.

\begin{figure*}
\includegraphics[width=1.5\columnwidth]{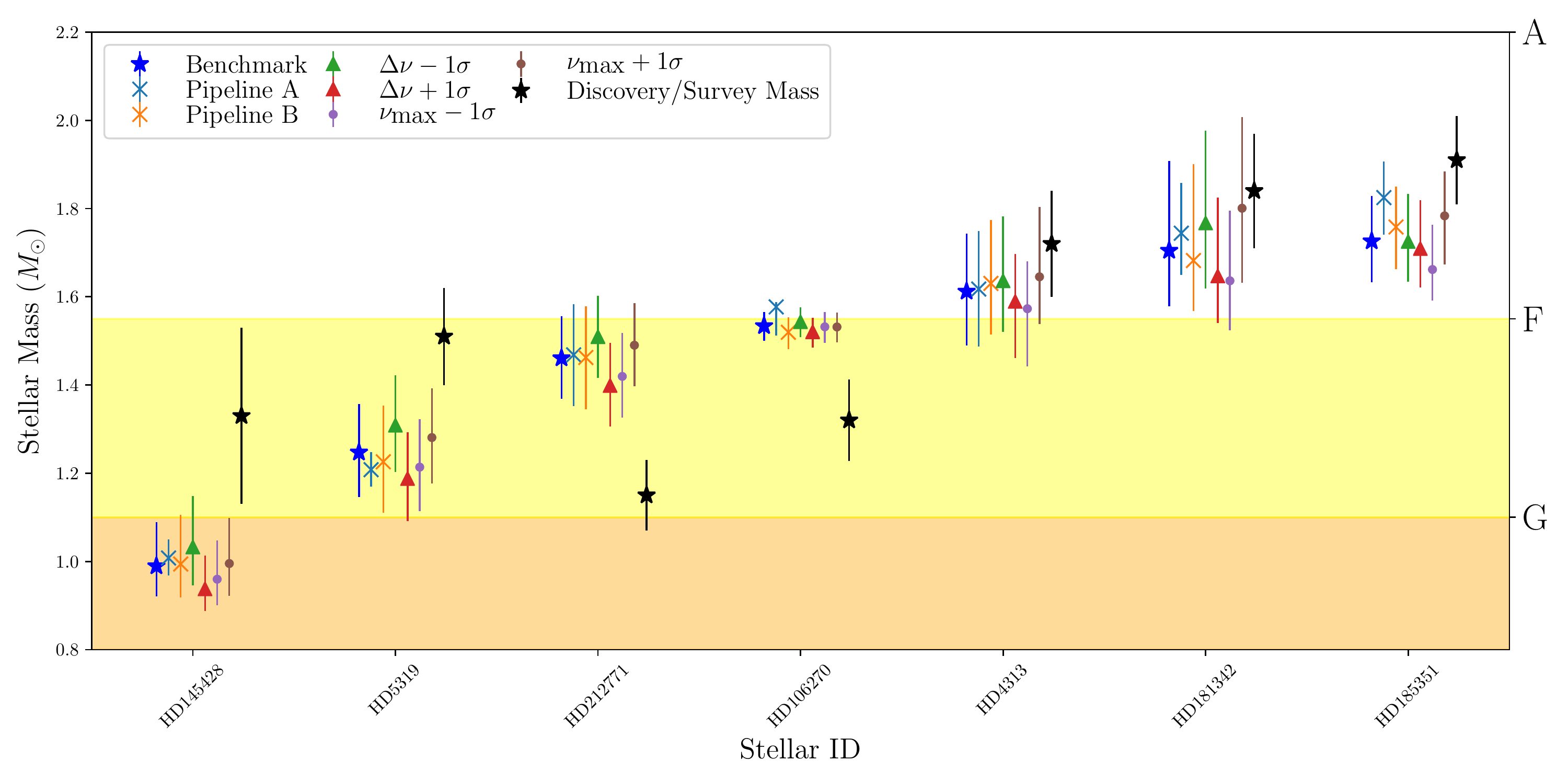}
\caption{Investigating the effect of biases in the seismic parameters, and potentially underestimated error bars. Black stars are the discovery masses, blue stars are the benchmark asteroseismic masses in Table \ref{tab:res}. Pipeline A and B (crosses) are two of the pipeline used to recover the asteroseismic parameters. Triangles are the $\Delta\nu$ values in Table \ref{tab:res} perturbed by $1\sigma$. Dots are the $\nu_{\textrm{max}}$ values in Table \ref{tab:res} perturbed by $1\sigma$.}
\label{fig:seis_bias}
\end{figure*}

In Sec \ref{sec:anal} we added additional uncertainties to the error bars on the seismic quantities returned by the \cite{Huber2009} pipeline to account for scatter between pipelines. Here, for completeness, we also tested using as inputs the seismic parameters and formal uncertainties from the other two pipelines. Again, we found very small changes in mass (at the level of or smaller than the uncertainties on the data). These results are shown in Fig \ref{fig:seis_bias}.

\section{Conclusions}\label{sec:conc}
This work has explored the masses of so-called ``retired A Stars'' and the impact of differing stellar models and the individual constraints used on the recovery of the stellar mass for single stars. In our ensemble of 7 stars, we find for 5 of the stars a mild shift to lower mass, when the asteroseismic mass is compared to the mass reported in the planet discovery paper. This mass shift is not significant. Additionally, the scale and sign of this mass offset is highly dependent on the chosen reference masses, as different literature masses for the ensemble cover the mass range $1-2\textrm{M}_{\odot}$, with optimistic error bars on literature masses resulting in significant offsets between different reference masses. We note that Stello et al. (in prep) find a similar, non-significant offset for stars of comparable mass to ours (from analysis of ground-based asteroseismic data collected on a sample of very bright A-type hosts), with evidence for an offset in a higher range of mass not explored in our sample. Stello et al. also find that the scatter on the literature values is of comparable size to the observed mass offset.

We also find that the mass difference can be explained through use of differing constraints during the recovery process. We also find that $\approx0.2\textrm{M}_{\odot}$ shifts in mass can be produced by only $1\sigma$ changes in temperature or metallicity, if only using spectroscopic and luminosity constraints. Additionally we find that even with the inclusion of asteroseismology, potentially inaccurate effective temperatures quoted with high precision makes the recovery of the true mass impossible. To solve this effective spectroscopic temperatures need calibrating to results from interferometry. Finally, we find that the use of optimistic uncertainties on input parameters has the potential to significantly bias the recovered stellar masses. The consequence of such an action would be to bias inferred planet occurrence rates, an argument broadly in agreement with the space-motion based argument of \cite{2013Schlaufman}, i.e., that the masses of evolved exoplanet hosts must be overestimated to explain the observed space motions of the same stars. 

Additionally an exploration of differences in recovered mass using differing stellar grids needs to be applied to a far larger number of stars, along with a full exploration of which underlying physical parameters are the cause of systematic shifts in mass. This will be the product of future work. 
Asteroseismic observations of more evolved exoplanet hosts will also be provided by data from later K2 campaigns and from the upcoming NASA TESS Mission (e.g. see \citealt{Campante2016}).

\section*{Acknowledgements}
The authors would like to thank D. Stello for helpful discussion. We also wish to thank the anonymous referee for helpful comments. This paper includes data collected by the \emph{Kepler} and K2 mission. The authors acknowledge the support of the UK Science and Technology Facilities Council (STFC). Funding for the Stellar Astrophysics Centre is provided by the Danish National Research Foundation (Grant DNRF106). MNL acknowledges the support of The Danish Council for Independent Research | Natural Science (Grant DFF-4181-00415). This research has made use of the SIMBAD database, operated at CDS, Strasbourg, France. 




\bibliographystyle{mnras}
\bibliography{biblio.bib} 

\begin{thebibliography}{}
\makeatletter
\relax
\def\mn@urlcharsother{\let\do\@makeother \do\$\do\&\do\#\do\^\do\_\do\%\do\~}
\def\mn@doi{\begingroup\mn@urlcharsother \@ifnextchar [ {\mn@doi@}
  {\mn@doi@[]}}
\def\mn@doi@[#1]#2{\def\@tempa{#1}\ifx\@tempa\@empty \href
  {http://dx.doi.org/#2} {doi:#2}\else \href {http://dx.doi.org/#2} {#1}\fi
  \endgroup}
\def\mn@eprint#1#2{\mn@eprint@#1:#2::\@nil}
\def\mn@eprint@arXiv#1{\href {http://arxiv.org/abs/#1} {{\tt arXiv:#1}}}
\def\mn@eprint@dblp#1{\href {http://dblp.uni-trier.de/rec/bibtex/#1.xml}
  {dblp:#1}}
\def\mn@eprint@#1:#2:#3:#4\@nil{\def\@tempa {#1}\def\@tempb {#2}\def\@tempc
  {#3}\ifx \@tempc \@empty \let \@tempc \@tempb \let \@tempb \@tempa \fi \ifx
  \@tempb \@empty \def\@tempb {arXiv}\fi \@ifundefined
  {mn@eprint@\@tempb}{\@tempb:\@tempc}{\expandafter \expandafter \csname
  mn@eprint@\@tempb\endcsname \expandafter{\@tempc}}}

\bibitem[\protect\citeauthoryear{{Akeson} et~al.,}{{Akeson}
  et~al.}{2013}]{ExArc2013}
{Akeson} R.~L.,  et~al., 2013, \mn@doi [\pasp] {10.1086/672273}, \href
  {http://adsabs.harvard.edu/abs/2013PASP..125..989A} {125, 989}

\bibitem[\protect\citeauthoryear{{Angulo} et~al.,}{{Angulo}
  et~al.}{1999}]{1999Angulo}
{Angulo} C.,  et~al., 1999, \mn@doi [Nuclear Physics A]
  {10.1016/S0375-9474(99)00030-5}, \href
  {http://adsabs.harvard.edu/abs/1999NuPhA.656....3A} {656, 3}

\bibitem[\protect\citeauthoryear{{Beck} et~al.,}{{Beck}
  et~al.}{2011}]{2011Beck}
{Beck} P.~G.,  et~al., 2011, \mn@doi [Science] {10.1126/science.1201939}, \href
  {http://adsabs.harvard.edu/abs/2011Sci...332..205B} {332, 205}

\bibitem[\protect\citeauthoryear{{Becker}, {Vanderburg}, {Adams}, {Rappaport}
  \& {Schwengeler}}{{Becker} et~al.}{2015}]{2015Becker}
{Becker} J.~C.,  {Vanderburg} A.,  {Adams} F.~C.,  {Rappaport} S.~A.,
  {Schwengeler} H.~M.,  2015, \mn@doi [\apjl] {10.1088/2041-8205/812/2/L18},
  \href {http://adsabs.harvard.edu/abs/2015ApJ...812L..18B} {812, L18}

\bibitem[\protect\citeauthoryear{{Bedding} et~al.,}{{Bedding}
  et~al.}{2011}]{2011Bedding}
{Bedding} T.~R.,  et~al., 2011, \mn@doi [\nat] {10.1038/nature09935}, \href
  {http://adsabs.harvard.edu/abs/2011Natur.471..608B} {471, 608}

\bibitem[\protect\citeauthoryear{{Blanco-Cuaresma} et~al.,}{{Blanco-Cuaresma}
  et~al.}{2016}]{2016Blanco}
{Blanco-Cuaresma} S.,  et~al., 2016, in 19th Cambridge Workshop on Cool Stars,
  Stellar Systems, and the Sun (CS19). p.~22 (\mn@eprint {arXiv} {1609.08092}),
  \mn@doi{10.5281/zenodo.155115}

\bibitem[\protect\citeauthoryear{{Bonfanti}, {Ortolani}, {Piotto}  \&
  {Nascimbeni}}{{Bonfanti} et~al.}{2015}]{2015Bofanti}
{Bonfanti} A.,  {Ortolani} S.,  {Piotto} G.,   {Nascimbeni} V.,  2015, \mn@doi
  [\aap] {10.1051/0004-6361/201424951}, \href
  {http://cdsads.u-strasbg.fr/abs/2015A%26A...575A..18B} {575, A18}

\bibitem[\protect\citeauthoryear{{Bonfanti}, {Ortolani}  \&
  {Nascimbeni}}{{Bonfanti} et~al.}{2016}]{2016Bofanti}
{Bonfanti} A.,  {Ortolani} S.,   {Nascimbeni} V.,  2016, \mn@doi [\aap]
  {10.1051/0004-6361/201527297}, \href
  {http://cdsads.u-strasbg.fr/abs/2016A%26A...585A...5B} {585, A5}

\bibitem[\protect\citeauthoryear{{Borucki} et~al.,}{{Borucki}
  et~al.}{2010}]{Borucki2010}
{Borucki} W.~J.,  et~al., 2010, \mn@doi [Science] {10.1126/science.1185402},
  \href {http://adsabs.harvard.edu/abs/2010Sci...327..977B} {327, 977}

\bibitem[\protect\citeauthoryear{{Bossini} et~al.,}{{Bossini}
  et~al.}{2015}]{2015Bossini}
{Bossini} D.,  et~al., 2015, \mn@doi [\mnras] {10.1093/mnras/stv1738}, \href
  {http://adsabs.harvard.edu/abs/2015MNRAS.453.2290B} {453, 2290}

\bibitem[\protect\citeauthoryear{{Bovy}, {Rix}, {Green}, {Schlafly}  \&
  {Finkbeiner}}{{Bovy} et~al.}{2016}]{mwdust}
{Bovy} J.,  {Rix} H.-W.,  {Green} G.~M.,  {Schlafly} E.~F.,   {Finkbeiner}
  D.~P.,  2016, \mn@doi [\apj] {10.3847/0004-637X/818/2/130}, \href
  {http://adsabs.harvard.edu/abs/2016ApJ...818..130B} {818, 130}

\bibitem[\protect\citeauthoryear{{Bowler} et~al.,}{{Bowler}
  et~al.}{2010}]{Bowler2010}
{Bowler} B.~P.,  et~al., 2010, \mn@doi [\apj] {10.1088/0004-637X/709/1/396},
  \href {http://adsabs.harvard.edu/abs/2010ApJ...709..396B} {709, 396}

\bibitem[\protect\citeauthoryear{{Bressan}, {Marigo}, {Girardi}, {Salasnich},
  {Dal Cero}, {Rubele}  \& {Nanni}}{{Bressan} et~al.}{2012}]{2012Bressan}
{Bressan} A.,  {Marigo} P.,  {Girardi} L.,  {Salasnich} B.,  {Dal Cero} C.,
  {Rubele} S.,   {Nanni} A.,  2012, \mn@doi [\mnras]
  {10.1111/j.1365-2966.2012.21948.x}, \href
  {http://adsabs.harvard.edu/abs/2012MNRAS.427..127B} {427, 127}

\bibitem[\protect\citeauthoryear{{Campante} et~al.,}{{Campante}
  et~al.}{2016}]{Campante2016}
{Campante} T.~L.,  et~al., 2016, \mn@doi [\apj] {10.3847/0004-637X/830/2/138},
  \href {http://adsabs.harvard.edu/abs/2016ApJ...830..138C} {830, 138}

\bibitem[\protect\citeauthoryear{{Campante} et~al.,}{{Campante}
  et~al.}{2017}]{2017arXiv170401794C}
{Campante} T.~L.,  et~al., 2017, preprint, \href
  {http://adsabs.harvard.edu/abs/2017arXiv170401794C} {} (\mn@eprint {arXiv}
  {1704.01794})

\bibitem[\protect\citeauthoryear{{Chaplin} \& {Miglio}}{{Chaplin} \&
  {Miglio}}{2013}]{Chaplin2013}
{Chaplin} W.~J.,  {Miglio} A.,  2013, \mn@doi [\araa]
  {10.1146/annurev-astro-082812-140938}, \href
  {http://adsabs.harvard.edu/abs/2013ARA%26A..51..353C} {51, 353}

\bibitem[\protect\citeauthoryear{{Chaplin} et~al.,}{{Chaplin}
  et~al.}{2015}]{Chaplin2015_K2C1}
{Chaplin} W.~J.,  et~al., 2015, \mn@doi [\pasp] {10.1086/683103}, \href
  {http://adsabs.harvard.edu/abs/2015PASP..127.1038C} {127, 1038}

\bibitem[\protect\citeauthoryear{{Choi}, {Dotter}, {Conroy}, {Cantiello},
  {Paxton}  \& {Johnson}}{{Choi} et~al.}{2016}]{2016Choi}
{Choi} J.,  {Dotter} A.,  {Conroy} C.,  {Cantiello} M.,  {Paxton} B.,
  {Johnson} B.~D.,  2016, \mn@doi [\apj] {10.3847/0004-637X/823/2/102}, \href
  {http://adsabs.harvard.edu/abs/2016ApJ...823..102C} {823, 102}

\bibitem[\protect\citeauthoryear{{Davies} \& {Miglio}}{{Davies} \&
  {Miglio}}{2016}]{2016Davies}
{Davies} G.~R.,  {Miglio} A.,  2016, \mn@doi [Astronomische Nachrichten]
  {10.1002/asna.201612371}, \href
  {http://adsabs.harvard.edu/abs/2016AN....337..774D} {337, 774}

\bibitem[\protect\citeauthoryear{{Davies}, {Handberg}, {Miglio}, {Campante},
  {Chaplin}  \& {Elsworth}}{{Davies} et~al.}{2014}]{2014Davies}
{Davies} G.~R.,  {Handberg} R.,  {Miglio} A.,  {Campante} T.~L.,  {Chaplin}
  W.~J.,   {Elsworth} Y.,  2014, \mn@doi [\mnras] {10.1093/mnrasl/slu143},
  \href {http://adsabs.harvard.edu/abs/2014MNRAS.445L..94D} {445, L94}

\bibitem[\protect\citeauthoryear{{Davies} et~al.,}{{Davies}
  et~al.}{2017}]{2017davies}
{Davies} G.~R.,  et~al., 2017, \mn@doi [\aap] {10.1051/0004-6361/201630066},
  \href {http://adsabs.harvard.edu/abs/2017A%26A...598L...4D} {598, L4}

\bibitem[\protect\citeauthoryear{{De Ridder}, {Molenberghs}, {Eyer}  \&
  {Aerts}}{{De Ridder} et~al.}{2016}]{2016deridder}
{De Ridder} J.,  {Molenberghs} G.,  {Eyer} L.,   {Aerts} C.,  2016, \mn@doi
  [\aap] {10.1051/0004-6361/201629799}, \href
  {http://adsabs.harvard.edu/abs/2016A%26A...595L...3D} {595, L3}

\bibitem[\protect\citeauthoryear{{Ferguson}, {Alexander}, {Allard}, {Barman},
  {Bodnarik}, {Hauschildt}, {Heffner-Wong}  \& {Tamanai}}{{Ferguson}
  et~al.}{2005}]{2005Ferguson}
{Ferguson} J.~W.,  {Alexander} D.~R.,  {Allard} F.,  {Barman} T.,  {Bodnarik}
  J.~G.,  {Hauschildt} P.~H.,  {Heffner-Wong} A.,   {Tamanai} A.,  2005,
  \mn@doi [\apj] {10.1086/428642}, \href
  {http://adsabs.harvard.edu/abs/2005ApJ...623..585F} {623, 585}

\bibitem[\protect\citeauthoryear{{Fischer} \& {Valenti}}{{Fischer} \&
  {Valenti}}{2005}]{2005Fish}
{Fischer} D.~A.,  {Valenti} J.,  2005, \mn@doi [\apj] {10.1086/428383}, \href
  {http://adsabs.harvard.edu/abs/2005ApJ...622.1102F} {622, 1102}

\bibitem[\protect\citeauthoryear{{Flower}}{{Flower}}{1996}]{Flowers1996}
{Flower} P.~J.,  1996, \mn@doi [\apj] {10.1086/177785}, \href
  {http://adsabs.harvard.edu/abs/1996ApJ...469..355F} {469, 355}

\bibitem[\protect\citeauthoryear{{Fressin} et~al.,}{{Fressin}
  et~al.}{2013}]{2013Fressin}
{Fressin} F.,  et~al., 2013, \mn@doi [\apj] {10.1088/0004-637X/766/2/81}, \href
  {http://adsabs.harvard.edu/abs/2013ApJ...766...81F} {766, 81}

\bibitem[\protect\citeauthoryear{{Fuller}, {Cantiello}, {Stello}, {Garcia}  \&
  {Bildsten}}{{Fuller} et~al.}{2015}]{2015Fuller}
{Fuller} J.,  {Cantiello} M.,  {Stello} D.,  {Garcia} R.~A.,   {Bildsten} L.,
  2015, \mn@doi [Science] {10.1126/science.aac6933}, \href
  {http://adsabs.harvard.edu/abs/2015Sci...350..423F} {350, 423}

\bibitem[\protect\citeauthoryear{{Ghezzi} \& {Johnson}}{{Ghezzi} \&
  {Johnson}}{2015}]{2015Ghezzi}
{Ghezzi} L.,  {Johnson} J.~A.,  2015, \mn@doi [\apj]
  {10.1088/0004-637X/812/2/96}, \href
  {http://adsabs.harvard.edu/abs/2015ApJ...812...96G} {812, 96}

\bibitem[\protect\citeauthoryear{{Ghezzi}, {do Nascimento}  \&
  {Johnson}}{{Ghezzi} et~al.}{2015}]{HD185351_2015}
{Ghezzi} L.,  {do Nascimento} Jr. J.-D.,   {Johnson} J.~A.,  2015, in {van
  Belle} G.~T.,  {Harris} H.~C.,  eds,  Cambridge Workshop on Cool Stars,
  Stellar Systems, and the Sun Vol. 18, 18th Cambridge Workshop on Cool Stars,
  Stellar Systems, and the Sun. pp 743--748 (\mn@eprint {arXiv} {1408.3143})

\bibitem[\protect\citeauthoryear{{Giguere}, {Fischer}, {Payne}, {Brewer},
  {Johnson}, {Howard}  \& {Isaacson}}{{Giguere} et~al.}{2015}]{2015Gig}
{Giguere} M.~J.,  {Fischer} D.~A.,  {Payne} M.~J.,  {Brewer} J.~M.,  {Johnson}
  J.~A.,  {Howard} A.~W.,   {Isaacson} H.~T.,  2015, \mn@doi [\apj]
  {10.1088/0004-637X/799/1/89}, \href
  {http://adsabs.harvard.edu/abs/2015ApJ...799...89G} {799, 89}

\bibitem[\protect\citeauthoryear{{Gilliland} et~al.,}{{Gilliland}
  et~al.}{2010}]{2010Gill}
{Gilliland} R.~L.,  et~al., 2010, \mn@doi [\pasp] {10.1086/650399}, \href
  {http://adsabs.harvard.edu/abs/2010PASP..122..131G} {122, 131}

\bibitem[\protect\citeauthoryear{{Green} et~al.,}{{Green}
  et~al.}{2015}]{Green2015}
{Green} G.~M.,  et~al., 2015, \mn@doi [\apj] {10.1088/0004-637X/810/1/25},
  \href {http://adsabs.harvard.edu/abs/2015ApJ...810...25G} {810, 25}

\bibitem[\protect\citeauthoryear{{Grevesse} \& {Noels}}{{Grevesse} \&
  {Noels}}{1993}]{1993Grevesse}
{Grevesse} N.,  {Noels} A.,  1993, \mn@doi [Physica Scripta Volume T]
  {10.1088/0031-8949/1993/T47/021}, \href
  {http://adsabs.harvard.edu/abs/1993PhST...47..133G} {47, 133}

\bibitem[\protect\citeauthoryear{{Handberg} \& {Lund}}{{Handberg} \&
  {Lund}}{2014}]{kasocfilt}
{Handberg} R.,  {Lund} M.~N.,  2014, \mn@doi [\mnras] {10.1093/mnras/stu1823},
  \href {http://adsabs.harvard.edu/abs/2014MNRAS.445.2698H} {445, 2698}

\bibitem[\protect\citeauthoryear{{Hekker} \& {Mel{\'e}ndez}}{{Hekker} \&
  {Mel{\'e}ndez}}{2007}]{2007Hekker}
{Hekker} S.,  {Mel{\'e}ndez} J.,  2007, \mn@doi [\aap]
  {10.1051/0004-6361:20078233}, \href
  {http://adsabs.harvard.edu/abs/2007A%26A...475.1003H} {475, 1003}

\bibitem[\protect\citeauthoryear{{Hekker} et~al.,}{{Hekker}
  et~al.}{2011}]{2011Hekker}
{Hekker} S.,  et~al., 2011, \mn@doi [\mnras]
  {10.1111/j.1365-2966.2011.18574.x}, \href
  {http://adsabs.harvard.edu/abs/2011MNRAS.414.2594H} {414, 2594}

\bibitem[\protect\citeauthoryear{{Hj{\o}rringgaard}, {Silva Aguirre}, {White},
  {Huber}, {Pope}, {Casagrande}, {Justesen}  \&
  {Christensen-Dalsgaard}}{{Hj{\o}rringgaard} et~al.}{2016}]{HD185351_2016}
{Hj{\o}rringgaard} J.~G.,  {Silva Aguirre} V.,  {White} T.~R.,  {Huber} D.,
  {Pope} B.~J.~S.,  {Casagrande} L.,  {Justesen} A.~B.,
  {Christensen-Dalsgaard} J.,  2016, \mn@doi [\mnras] {10.1093/mnras/stw2559},
  \href {http://adsabs.harvard.edu/abs/2016MNRAS.tmp.1542H} {}

\bibitem[\protect\citeauthoryear{{Howard} et~al.,}{{Howard}
  et~al.}{2010}]{2010Howard}
{Howard} A.~W.,  et~al., 2010, \mn@doi [Science] {10.1126/science.1194854},
  \href {http://adsabs.harvard.edu/abs/2010Sci...330..653H} {330, 653}

\bibitem[\protect\citeauthoryear{{Howell} et~al.,}{{Howell}
  et~al.}{2014}]{Howell2014}
{Howell} S.~B.,  et~al., 2014, \mn@doi [\pasp] {10.1086/676406}, \href
  {http://adsabs.harvard.edu/abs/2014PASP..126..398H} {126, 398}

\bibitem[\protect\citeauthoryear{{Huber}, {Stello}, {Bedding}, {Chaplin},
  {Arentoft}, {Quirion}  \& {Kjeldsen}}{{Huber} et~al.}{2009}]{Huber2009}
{Huber} D.,  {Stello} D.,  {Bedding} T.~R.,  {Chaplin} W.~J.,  {Arentoft} T.,
  {Quirion} P.-O.,   {Kjeldsen} H.,  2009, Communications in Asteroseismology,
  \href {http://adsabs.harvard.edu/abs/2009CoAst.160...74H} {160, 74}

\bibitem[\protect\citeauthoryear{{Huber} et~al.,}{{Huber}
  et~al.}{2010}]{2010Huber}
{Huber} D.,  et~al., 2010, \mn@doi [\apj] {10.1088/0004-637X/723/2/1607}, \href
  {http://adsabs.harvard.edu/abs/2010ApJ...723.1607H} {723, 1607}

\bibitem[\protect\citeauthoryear{{Huber} et~al.,}{{Huber}
  et~al.}{2013}]{Huber2013}
{Huber} D.,  et~al., 2013, \mn@doi [\apj] {10.1088/0004-637X/767/2/127}, \href
  {http://adsabs.harvard.edu/abs/2013ApJ...767..127H} {767, 127}

\bibitem[\protect\citeauthoryear{{Huber} et~al.,}{{Huber}
  et~al.}{2014}]{Huber2014}
{Huber} D.,  et~al., 2014, \mn@doi [\apjs] {10.1088/0067-0049/211/1/2}, \href
  {http://adsabs.harvard.edu/abs/2014ApJS..211....2H} {211, 2}

\bibitem[\protect\citeauthoryear{{Huber} et~al.,}{{Huber}
  et~al.}{2016}]{2016Huber}
{Huber} D.,  et~al., 2016, \mn@doi [\apjs] {10.3847/0067-0049/224/1/2}, \href
  {http://adsabs.harvard.edu/abs/2016ApJS..224....2H} {224, 2}

\bibitem[\protect\citeauthoryear{{Huber} et~al.,}{{Huber}
  et~al.}{2017}]{2017arXiv170504697H}
{Huber} D.,  et~al., 2017, preprint, \href
  {http://adsabs.harvard.edu/abs/2017arXiv170504697H} {} (\mn@eprint {arXiv}
  {1705.04697})

\bibitem[\protect\citeauthoryear{{Iglesias} \& {Rogers}}{{Iglesias} \&
  {Rogers}}{1996}]{1996Iglesias}
{Iglesias} C.~A.,  {Rogers} F.~J.,  1996, \mn@doi [\apj] {10.1086/177381},
  \href {http://adsabs.harvard.edu/abs/1996ApJ...464..943I} {464, 943}

\bibitem[\protect\citeauthoryear{{Jofr{\'e}}, {Petrucci}, {Saffe}, {Saker}, {de
  la Villarmois}, {Chavero}, {G{\'o}mez}  \& {Mauas}}{{Jofr{\'e}}
  et~al.}{2015}]{Jofre2015}
{Jofr{\'e}} E.,  {Petrucci} R.,  {Saffe} C.,  {Saker} L.,  {de la Villarmois}
  E.~A.,  {Chavero} C.,  {G{\'o}mez} M.,   {Mauas} P.~J.~D.,  2015, \mn@doi
  [\aap] {10.1051/0004-6361/201424474}, \href
  {http://adsabs.harvard.edu/abs/2015A%26A...574A..50J} {574, A50}

\bibitem[\protect\citeauthoryear{{Johnson}, {Marcy}, {Fischer}, {Henry},
  {Wright}, {Isaacson}  \& {McCarthy}}{{Johnson} et~al.}{2006}]{Johnson2006}
{Johnson} J.~A.,  {Marcy} G.~W.,  {Fischer} D.~A.,  {Henry} G.~W.,  {Wright}
  J.~T.,  {Isaacson} H.,   {McCarthy} C.,  2006, \mn@doi [\apj]
  {10.1086/508255}, \href {http://adsabs.harvard.edu/abs/2006ApJ...652.1724J}
  {652, 1724}

\bibitem[\protect\citeauthoryear{{Johnson} et~al.,}{{Johnson}
  et~al.}{2007a}]{Johnson2007a}
{Johnson} J.~A.,  et~al., 2007a, \mn@doi [\apj] {10.1086/519677}, \href
  {http://adsabs.harvard.edu/abs/2007ApJ...665..785J} {665, 785}

\bibitem[\protect\citeauthoryear{{Johnson}, {Butler}, {Marcy}, {Fischer},
  {Vogt}, {Wright}  \& {Peek}}{{Johnson} et~al.}{2007b}]{Johnson2007b}
{Johnson} J.~A.,  {Butler} R.~P.,  {Marcy} G.~W.,  {Fischer} D.~A.,  {Vogt}
  S.~S.,  {Wright} J.~T.,   {Peek} K.~M.~G.,  2007b, \mn@doi [\apj]
  {10.1086/521720}, \href {http://adsabs.harvard.edu/abs/2007ApJ...670..833J}
  {670, 833}

\bibitem[\protect\citeauthoryear{{Johnson}, {Marcy}, {Fischer}, {Wright},
  {Reffert}, {Kregenow}, {Williams}  \& {Peek}}{{Johnson}
  et~al.}{2008}]{Johnson2008}
{Johnson} J.~A.,  {Marcy} G.~W.,  {Fischer} D.~A.,  {Wright} J.~T.,  {Reffert}
  S.,  {Kregenow} J.~M.,  {Williams} P.~K.~G.,   {Peek} K.~M.~G.,  2008,
  \mn@doi [\apj] {10.1086/526453}, \href
  {http://adsabs.harvard.edu/abs/2008ApJ...675..784J} {675, 784}

\bibitem[\protect\citeauthoryear{{Johnson}, {Howard}, {Bowler}, {Henry},
  {Marcy}, {Wright}, {Fischer}  \& {Isaacson}}{{Johnson}
  et~al.}{2010a}]{johnson2010b}
{Johnson} J.~A.,  {Howard} A.~W.,  {Bowler} B.~P.,  {Henry} G.~W.,  {Marcy}
  G.~W.,  {Wright} J.~T.,  {Fischer} D.~A.,   {Isaacson} H.,  2010a, \mn@doi
  [\pasp] {10.1086/653809}, \href
  {http://adsabs.harvard.edu/abs/2010PASP..122..701J} {122, 701}

\bibitem[\protect\citeauthoryear{{Johnson}, {Aller}, {Howard}  \&
  {Crepp}}{{Johnson} et~al.}{2010b}]{Johnson2010}
{Johnson} J.~A.,  {Aller} K.~M.,  {Howard} A.~W.,   {Crepp} J.~R.,  2010b,
  \mn@doi [\pasp] {10.1086/655775}, \href
  {http://adsabs.harvard.edu/abs/2010PASP..122..905J} {122, 905}

\bibitem[\protect\citeauthoryear{{Johnson} et~al.,}{{Johnson}
  et~al.}{2011}]{2011Johnson}
{Johnson} J.~A.,  et~al., 2011, \mn@doi [\apjs] {10.1088/0067-0049/197/2/26},
  \href {http://adsabs.harvard.edu/abs/2011ApJS..197...26J} {197, 26}

\bibitem[\protect\citeauthoryear{{Johnson} et~al.,}{{Johnson}
  et~al.}{2014}]{2014JohnsonHuber}
{Johnson} J.~A.,  et~al., 2014, \mn@doi [\apj] {10.1088/0004-637X/794/1/15},
  \href {http://adsabs.harvard.edu/abs/2014ApJ...794...15J} {794, 15}

\bibitem[\protect\citeauthoryear{{Jones} et~al.,}{{Jones}
  et~al.}{2016}]{2016MJones}
{Jones} M.~I.,  et~al., 2016, \mn@doi [\aap] {10.1051/0004-6361/201628067},
  \href {http://adsabs.harvard.edu/abs/2016A%26A...590A..38J} {590, A38}

\bibitem[\protect\citeauthoryear{{Krishna Swamy}}{{Krishna
  Swamy}}{1966}]{1966Kris}
{Krishna Swamy} K.~S.,  1966, \mn@doi [\apj] {10.1086/148752}, \href
  {http://adsabs.harvard.edu/abs/1966ApJ...145..174K} {145, 174}

\bibitem[\protect\citeauthoryear{{Lagarde}, {Bossini}, {Miglio}, {Vrard}  \&
  {Mosser}}{{Lagarde} et~al.}{2016}]{2016Lagarde}
{Lagarde} N.,  {Bossini} D.,  {Miglio} A.,  {Vrard} M.,   {Mosser} B.,  2016,
  \mn@doi [\mnras] {10.1093/mnrasl/slv201}, \href
  {http://adsabs.harvard.edu/abs/2016MNRAS.457L..59L} {457, L59}

\bibitem[\protect\citeauthoryear{{Lindegren} et~al.,}{{Lindegren}
  et~al.}{2016}]{GAIADR1}
{Lindegren} L.,  et~al., 2016, preprint, \href
  {http://adsabs.harvard.edu/abs/2016arXiv160904303L} {} (\mn@eprint {arXiv}
  {1609.04303})

\bibitem[\protect\citeauthoryear{{Lloyd}}{{Lloyd}}{2011}]{2011Lloyd}
{Lloyd} J.~P.,  2011, \mn@doi [\apjl] {10.1088/2041-8205/739/2/L49}, \href
  {http://adsabs.harvard.edu/abs/2011ApJ...739L..49L} {739, L49}

\bibitem[\protect\citeauthoryear{{Lund}, {Handberg}, {Davies}, {Chaplin}  \&
  {Jones}}{{Lund} et~al.}{2015}]{K2P2}
{Lund} M.~N.,  {Handberg} R.,  {Davies} G.~R.,  {Chaplin} W.~J.,   {Jones}
  C.~D.,  2015, \mn@doi [\apj] {10.1088/0004-637X/806/1/30}, \href
  {http://adsabs.harvard.edu/abs/2015ApJ...806...30L} {806, 30}

\bibitem[\protect\citeauthoryear{{Lund} et~al.,}{{Lund}
  et~al.}{2016}]{2016Lund_hyades}
{Lund} M.~N.,  et~al., 2016, \mn@doi [\mnras] {10.1093/mnras/stw2160}, \href
  {http://adsabs.harvard.edu/abs/2016MNRAS.463.2600L} {463, 2600}

\bibitem[\protect\citeauthoryear{{Maeder}}{{Maeder}}{1975}]{1975Maeder}
{Maeder} A.,  1975, \aap, \href
  {http://adsabs.harvard.edu/abs/1975A%26A....40..303M} {40, 303}

\bibitem[\protect\citeauthoryear{{Maldonado}, {Villaver}  \&
  {Eiroa}}{{Maldonado} et~al.}{2013}]{2013Maldo}
{Maldonado} J.,  {Villaver} E.,   {Eiroa} C.,  2013, \mn@doi [\aap]
  {10.1051/0004-6361/201321082}, \href
  {http://adsabs.harvard.edu/abs/2013A%26A...554A..84M} {554, A84}

\bibitem[\protect\citeauthoryear{{Mortier}, {Santos}, {Sousa}, {Adibekyan},
  {Delgado Mena}, {Tsantaki}, {Israelian}  \& {Mayor}}{{Mortier}
  et~al.}{2013}]{Mortier2013}
{Mortier} A.,  {Santos} N.~C.,  {Sousa} S.~G.,  {Adibekyan} V.~Z.,  {Delgado
  Mena} E.,  {Tsantaki} M.,  {Israelian} G.,   {Mayor} M.,  2013, \mn@doi
  [\aap] {10.1051/0004-6361/201321641}, \href
  {http://adsabs.harvard.edu/abs/2013A%26A...557A..70M} {557, A70}

\bibitem[\protect\citeauthoryear{{Mosser} et~al.,}{{Mosser}
  et~al.}{2011}]{2011MosserCorot}
{Mosser} B.,  et~al., 2011, \mn@doi [\aap] {10.1051/0004-6361/201116825}, \href
  {http://adsabs.harvard.edu/abs/2011A%26A...532A..86M} {532, A86}

\bibitem[\protect\citeauthoryear{{Mosser} et~al.,}{{Mosser}
  et~al.}{2012}]{Mosser2012}
{Mosser} B.,  et~al., 2012, \mn@doi [\aap] {10.1051/0004-6361/201117352}, \href
  {http://adsabs.harvard.edu/abs/2012A%26A...537A..30M} {537, A30}

\bibitem[\protect\citeauthoryear{{Mosser} et~al.,}{{Mosser}
  et~al.}{2016}]{2016arXivMosser_dipole}
{Mosser} B.,  et~al., 2016, preprint, \href
  {http://adsabs.harvard.edu/abs/2016arXiv161003872M} {} (\mn@eprint {arXiv}
  {1610.03872})

\bibitem[\protect\citeauthoryear{{Mullally}, {Barclay}  \&
  {Barentsen}}{{Mullally} et~al.}{2016}]{k2fov}
{Mullally} F.,  {Barclay} T.,   {Barentsen} G.,  2016, {K2fov: Field of view
  software for NASA's K2 mission}, Astrophysics Source Code Library (\mn@eprint
  {ascl} {1601.009})

\bibitem[\protect\citeauthoryear{{Paxton}, {Bildsten}, {Dotter}, {Herwig},
  {Lesaffre}  \& {Timmes}}{{Paxton} et~al.}{2011}]{2011Paxton}
{Paxton} B.,  {Bildsten} L.,  {Dotter} A.,  {Herwig} F.,  {Lesaffre} P.,
  {Timmes} F.,  2011, \mn@doi [\apjs] {10.1088/0067-0049/192/1/3}, \href
  {http://adsabs.harvard.edu/abs/2011ApJS..192....3P} {192, 3}

\bibitem[\protect\citeauthoryear{{Paxton} et~al.,}{{Paxton}
  et~al.}{2013}]{2013Paxton}
{Paxton} B.,  et~al., 2013, \mn@doi [\apjs] {10.1088/0067-0049/208/1/4}, \href
  {http://adsabs.harvard.edu/abs/2013ApJS..208....4P} {208, 4}

\bibitem[\protect\citeauthoryear{{Pijpers}}{{Pijpers}}{2003}]{pijpers2003}
{Pijpers} F.~P.,  2003, \mn@doi [\aap] {10.1051/0004-6361:20021839}, \href
  {http://adsabs.harvard.edu/abs/2003A%26A...400..241P} {400, 241}

\bibitem[\protect\citeauthoryear{{Reffert}, {Bergmann}, {Quirrenbach},
  {Trifonov}  \& {K{\"u}nstler}}{{Reffert} et~al.}{2015}]{Reffert2014}
{Reffert} S.,  {Bergmann} C.,  {Quirrenbach} A.,  {Trifonov} T.,
  {K{\"u}nstler} A.,  2015, \mn@doi [\aap] {10.1051/0004-6361/201322360}, \href
  {http://adsabs.harvard.edu/abs/2015A%26A...574A.116R} {574, A116}

\bibitem[\protect\citeauthoryear{{Robinson} et~al.,}{{Robinson}
  et~al.}{2007}]{2007Robinson}
{Robinson} S.~E.,  et~al., 2007, \mn@doi [\apj] {10.1086/522106}, \href
  {http://adsabs.harvard.edu/abs/2007ApJ...670.1391R} {670, 1391}

\bibitem[\protect\citeauthoryear{{Rodrigues} et~al.,}{{Rodrigues}
  et~al.}{2017}]{2017Rod}
{Rodrigues} T.~S.,  et~al., 2017, \mn@doi [\mnras] {10.1093/mnras/stx120},
  \href {http://adsabs.harvard.edu/abs/2017MNRAS.467.1433R} {467, 1433}

\bibitem[\protect\citeauthoryear{{Rogers} \& {Nayfonov}}{{Rogers} \&
  {Nayfonov}}{2002}]{2002Rogers}
{Rogers} F.~J.,  {Nayfonov} A.,  2002, \mn@doi [\apj] {10.1086/341894}, \href
  {http://adsabs.harvard.edu/abs/2002ApJ...576.1064R} {576, 1064}

\bibitem[\protect\citeauthoryear{{Schlaufman} \& {Winn}}{{Schlaufman} \&
  {Winn}}{2013}]{2013Schlaufman}
{Schlaufman} K.~C.,  {Winn} J.~N.,  2013, \mn@doi [\apj]
  {10.1088/0004-637X/772/2/143}, \href
  {http://adsabs.harvard.edu/abs/2013ApJ...772..143S} {772, 143}

\bibitem[\protect\citeauthoryear{{Sousa} et~al.,}{{Sousa}
  et~al.}{2008}]{2008Sousa}
{Sousa} S.~G.,  et~al., 2008, \mn@doi [\aap] {10.1051/0004-6361:200809698},
  \href {http://adsabs.harvard.edu/abs/2008A%26A...487..373S} {487, 373}

\bibitem[\protect\citeauthoryear{{Stassun} \& {Torres}}{{Stassun} \&
  {Torres}}{2016}]{2016stassun}
{Stassun} K.~G.,  {Torres} G.,  2016, preprint, \href
  {http://adsabs.harvard.edu/abs/2016arXiv160905390S} {} (\mn@eprint {arXiv}
  {1609.05390})

\bibitem[\protect\citeauthoryear{{Stello} et~al.,}{{Stello}
  et~al.}{2013}]{Stello2013}
{Stello} D.,  et~al., 2013, \mn@doi [\apjl] {10.1088/2041-8205/765/2/L41},
  \href {http://adsabs.harvard.edu/abs/2013ApJ...765L..41S} {765, L41}

\bibitem[\protect\citeauthoryear{{Stello} et~al.,}{{Stello}
  et~al.}{2015}]{2015Stello}
{Stello} D.,  et~al., 2015, \mn@doi [\apjl] {10.1088/2041-8205/809/1/L3}, \href
  {http://adsabs.harvard.edu/abs/2015ApJ...809L...3S} {809, L3}

\bibitem[\protect\citeauthoryear{{Stello}, {Cantiello}, {Fuller}, {Huber},
  {Garc{\'{\i}}a}, {Bedding}, {Bildsten}  \& {Silva Aguirre}}{{Stello}
  et~al.}{2016}]{2016NatureStello}
{Stello} D.,  {Cantiello} M.,  {Fuller} J.,  {Huber} D.,  {Garc{\'{\i}}a}
  R.~A.,  {Bedding} T.~R.,  {Bildsten} L.,   {Silva Aguirre} V.,  2016, \mn@doi
  [\nat] {10.1038/nature16171}, \href
  {http://adsabs.harvard.edu/abs/2016Natur.529..364S} {529, 364}

\bibitem[\protect\citeauthoryear{{Torres}}{{Torres}}{2010}]{Torres2010}
{Torres} G.,  2010, \mn@doi [\aj] {10.1088/0004-6256/140/5/1158}, \href
  {http://adsabs.harvard.edu/abs/2010AJ....140.1158T} {140, 1158}

\bibitem[\protect\citeauthoryear{{Torres}, {Fischer}, {Sozzetti}, {Buchhave},
  {Winn}, {Holman}  \& {Carter}}{{Torres} et~al.}{2012}]{Torres2012}
{Torres} G.,  {Fischer} D.~A.,  {Sozzetti} A.,  {Buchhave} L.~A.,  {Winn}
  J.~N.,  {Holman} M.~J.,   {Carter} J.~A.,  2012, \mn@doi [\apj]
  {10.1088/0004-637X/757/2/161}, \href
  {http://adsabs.harvard.edu/abs/2012ApJ...757..161T} {757, 161}

\bibitem[\protect\citeauthoryear{{Tsantaki}, {Sousa}, {Adibekyan}, {Santos},
  {Mortier}  \& {Israelian}}{{Tsantaki} et~al.}{2013}]{2013Tsantaki}
{Tsantaki} M.,  {Sousa} S.~G.,  {Adibekyan} V.~Z.,  {Santos} N.~C.,  {Mortier}
  A.,   {Israelian} G.,  2013, \mn@doi [\aap] {10.1051/0004-6361/201321103},
  \href {http://adsabs.harvard.edu/abs/2013A%26A...555A.150T} {555, A150}

\bibitem[\protect\citeauthoryear{{Valenti} \& {Piskunov}}{{Valenti} \&
  {Piskunov}}{1996}]{1996Valenti_SME}
{Valenti} J.~A.,  {Piskunov} N.,  1996, \aaps, \href
  {http://adsabs.harvard.edu/abs/1996A%26AS..118..595V} {118, 595}

\bibitem[\protect\citeauthoryear{{Verner} et~al.,}{{Verner}
  et~al.}{2011}]{Verner2011}
{Verner} G.~A.,  et~al., 2011, \mn@doi [\mnras]
  {10.1111/j.1365-2966.2011.18968.x}, \href
  {http://adsabs.harvard.edu/abs/2011MNRAS.415.3539V} {415, 3539}

\bibitem[\protect\citeauthoryear{{Wittenmyer}, {Endl}, {Wang}, {Johnson},
  {Tinney}  \& {O'Toole}}{{Wittenmyer} et~al.}{2011}]{2011Witten}
{Wittenmyer} R.~A.,  {Endl} M.,  {Wang} L.,  {Johnson} J.~A.,  {Tinney} C.~G.,
   {O'Toole} S.~J.,  2011, \mn@doi [\apj] {10.1088/0004-637X/743/2/184}, \href
  {http://adsabs.harvard.edu/abs/2011ApJ...743..184W} {743, 184}

\bibitem[\protect\citeauthoryear{{Wittenmyer}, {Liu}, {Wang}, {Casagrande},
  {Johnson}  \& {Tinney}}{{Wittenmyer} et~al.}{2016}]{2016Witten}
{Wittenmyer} R.~A.,  {Liu} F.,  {Wang} L.,  {Casagrande} L.,  {Johnson} J.~A.,
   {Tinney} C.~G.,  2016, \mn@doi [\aj] {10.3847/0004-6256/152/1/19}, \href
  {http://adsabs.harvard.edu/abs/2016AJ....152...19W} {152, 19}

\bibitem[\protect\citeauthoryear{{da Silva} et~al.,}{{da Silva}
  et~al.}{2006}]{2006dasilva}
{da Silva} L.,  et~al., 2006, \mn@doi [\aap] {10.1051/0004-6361:20065105},
  \href {http://adsabs.harvard.edu/abs/2006A%26A...458..609D} {458, 609}

\bibitem[\protect\citeauthoryear{{van Leeuwen}}{{van Leeuwen}}{2007}]{2007HIP}
{van Leeuwen} F.,  2007, \mn@doi [\aap] {10.1051/0004-6361:20078357}, \href
  {http://adsabs.harvard.edu/abs/2007A%26A...474..653V} {474, 653}

\makeatother
\end{thebibliography}

\appendix
\section{Available literature masses}

\begin{table*}
	\centering
	\caption{All available literature masses for each star in the ensemble. The values are primarily estimated from the observed spectroscopic parameters. The mass values from \protect\cite{Huber2014} are estimated from the \emph{Hipparcos} parallax.}
	\label{tab:litmass}
	\begin{threeparttable}
	\centering
	\begin{tabular}{lllllllllllllllll} 
	\hline
	EPIC/KIC&HD&Huber\tnote{1}&Johnson\tnote{2}&Mortier\tnote{3,a}&Mortier\tnote{3,b}&Bofanti\tnote{4}&Bofanti\tnote{3,b}&Jofr\'e\tnote{6}&Maldonado\tnote{7}\\
\hline
203514293&145428&$1.274^{+0.516}_{-0.413}$&&&&&&&\\
220548055&4313&$1.94^{+0.039}_{-0.849}$&$1.72\pm0.12$&$1.53\pm0.09$&$1.35\pm0.11$&$1.72\pm0.03$&$1.49\pm0.04$&$1.71\pm0.13$&\\
215745876&181342&$1.203^{+0.176}_{-0.246}$&$1.84\pm0.13$&$1.7\pm0.09$&$1.49\pm0.19$&$1.40\pm0.1$&$1.40\pm0.1$&$1.78\pm0.11$&$1.78\pm0.07$\\
220222356&5319&$1.232^{+0.178}_{-0.250}$&&$1.28\pm0.1$&$1.24\pm0.14$&$1.40\pm0.1$&$1.2\pm0.1$&&\\
8566020&185351&&&&&&&$1.82\pm0.05$&\\
205924248&212771&$1.173^{+0.154}_{-0.263}$&$1.15\pm0.08$&$1.51\pm0.08$&$1.22\pm0.08$&$1.40\pm0.1$&$1.45\pm0.02$&$1.60\pm0.13$&$1.63\pm0.1$\\
228737206&106270&$1.447^{+0.119}_{-0.119}$&$1.32\pm0.092$\tnote{2a}&$1.33\pm0.05$&$1.33\pm0.06$&$1.35\pm0.02$&$1.37\pm0.03$&&\\
\hline
EPIC/KIC&HD&Wittenmyer\tnote{8}&Huber\tnote{9}&Jones\tnote{10}&Robinson\tnote{11}&Giguere\tnote{12}&Reffert\tnote{13}&Johnson\tnote{14}&Ghezzi\tnote{15}\\
\hline
203514293&145428&$1.33\pm0.2$&&&&&&&\\
220548055&4313&&&&&&&&\\
215745876&181342&$1.42\pm0.2$&&$1.89\pm0.11$&&&&&\\
220222356&5319&&&&$1.56\pm0.18$&$1.51\pm0.11$&&&\\
8566020&185351&&$1.684^{+0.166}_{-0.499}$&&&&$1.73\pm0.15$&$1.60\pm0.08$\tnote{c}&$1.77\pm0.04$\\
&&&&&&&&$1.99\pm0.23$\tnote{d}&\\
&&&&&&&&$1.90\pm0.15$\tnote{e}&\\
&&&&&&&&$1.87\pm0.07$\tnote{f}&\\
205924248&212771&&&&&&&&\\
228737206&106270&&&&&&&&\\
	\hline
	\end{tabular}
	\begin{tablenotes}
	\item[a] \cite{2013Tsantaki} line list for the stars cooler than 5200 K, and the \cite{2008Sousa} line list for the hotter stars.
	\item[b] \cite{2007Hekker} line list
	\end{tablenotes}
	\begin{tablenotes}
	\item[1] \cite{2016Huber}.
	\item[2] \cite{johnson2010b}.
	\item[2a] \cite{2011Johnson}	
	\item[3] \cite{Mortier2013}.
	\item[4] \cite{2015Bofanti}.
	\item[5] \cite{2016Bofanti}.
	\item[6] \cite{Jofre2015}.
	\item[7] \cite{2013Maldo}.
	\item[8] \cite{2016Witten}.
	\item[9] \cite{Huber2014}.
	\item[10] \cite{2016MJones}.
	\item[11] \cite{2007Robinson}.
	\item[12] \cite{2015Gig}.
	\item[13] \cite{Reffert2014}.
	\item[14] \cite{2014JohnsonHuber}.
		\item[c] Interferometric radius, combined with asteroseismology.
		\item[d] Scaling relation, based on $\Delta\nu=15.4\pm0.2\mu$Hz, $\nu_{\textrm{max}}=229.8\pm6.0\mu$Hz.
		\item[e] BaSTI Grid fitting with asteroseismology and SME spectroscopy.
		\item[f] Grid fitting with only SME spectroscopy, iterated with Y$^{2}$ grids to a converged $\log_{10}{g}$.
	\item[15] \cite{HD185351_2015}.
	\end{tablenotes}
	\end{threeparttable}
\end{table*}

\begin{figure*}
\includegraphics[width=1.5\columnwidth]{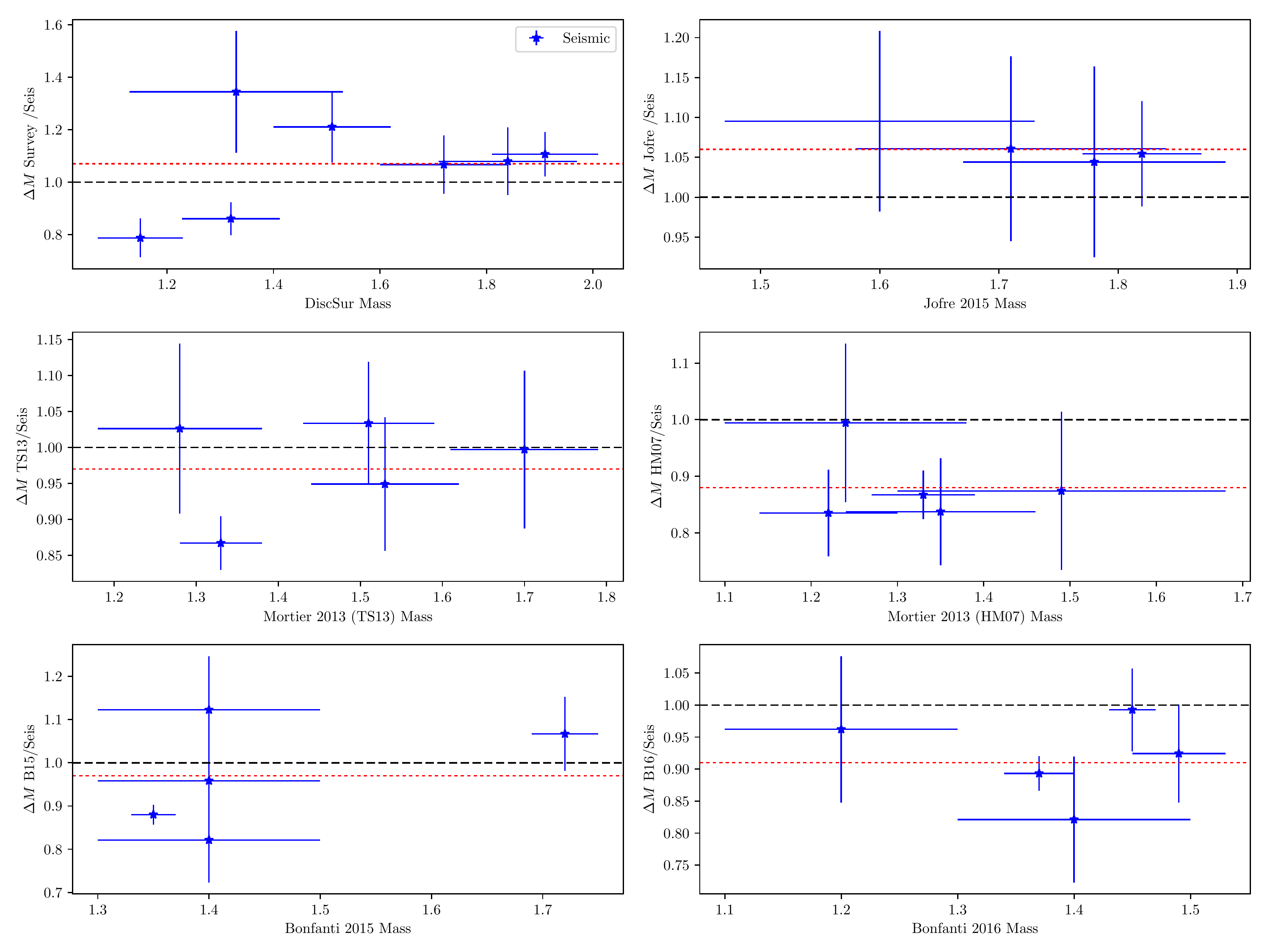}
\caption{For stars with several available literature masses, the ratio of seismic to literature mass is shown, against the literature mass value. The red dotted line in each subplot is the average ratio of Table \ref{tab:litcomp}. Black dashed line is parity}
\label{fig:massratio}
\end{figure*}

\bsp	
\label{lastpage}
\end{document}